\documentclass[final,3p,times,twocolumn]{elsarticle}
\usepackage{lineno}

\usepackage{graphicx}
\usepackage{subfig}
\usepackage{pfa}
\usepackage{mathptmx}
\usepackage{helvet}
\usepackage{amsmath}
\usepackage{color}
\usepackage{units}
\usepackage{url}
\usepackage{xspace}
\usepackage{booktabs}
\usepackage{multirow}
\usepackage{wrapfig}
\usepackage{upgreek}
\usepackage{threeparttable}
\usepackage{slashed}

\journal{Nucl. Instr. and Meth. A}

\begin{document}
  \begin{frontmatter}

    
    \title{Performance of Particle Flow Calorimetry at CLIC}
    
    \author[a]{J. S. Marshall}
    \author[b]{A. M\"unnich}
    \author[a]{M. A. Thomson}
    
    \address[a]{Cavendish Laboratory, University of Cambridge, Cambridge, United Kingdom}
    \address[b]{CERN, Geneva, Switzerland}
      
      \begin{abstract}
	The particle flow approach to calorimetry can provide unprecedented jet energy resolution at a future high energy collider, such as the International Linear Collider (ILC).
        However, the use of particle flow calorimetry at the proposed multi-TeV Compact Linear Collider (CLIC) poses a number of significant new challenges. At higher jet energies, detector
	occupancies increase, and it becomes increasingly difficult to resolve energy deposits from individual particles. The experimental conditions at CLIC are also significantly more 
	challenging than those at previous electron-positron colliders, with increased levels of beam-induced backgrounds combined with a bunch spacing of only 0.5\,ns. This paper describes the modifications
	made to the PandoraPFA particle flow algorithm to improve the jet energy reconstruction for jet energies above 250\,GeV. It then introduces a combination of timing and $p_{T}$ cuts that can be 
	applied to reconstructed particles in order to significantly reduce the background. A systematic study is performed to understand the dependence of the jet energy resolution on the jet energy
	and angle, and the physics performance is assessed via a study of the energy and mass resolution of $\Wboson$ and $\Zzero$ particles in the presence of background at CLIC. Finally, the missing 
	transverse momentum resolution is presented, and the fake missing momentum is quantified.
	The results presented in this paper demonstrate that high granularity particle flow calorimetry 
	 leads to a robust and high resolution reconstruction of jet energies and di-jet masses at CLIC. 
      \end{abstract}
      
  \end{frontmatter}
  

  \section{Introduction}
  CLIC~\cite{CLIC-CDR} is a proposed linear collider designed to perform electron-positron collisions at centre-of-mass energies ranging from a few hundred GeV up to 3\,TeV, with a luminosity of 
  $5.9\cdot10^{34}\,\mathrm{cm}^{-2}\,\mathrm{s}^{-1}$ at the highest energy. The anticipated physics potential of such a collider is extremely broad. It ranges from precision tests of the Higgs and top sector, 
  to detailed studies of new phenomena, through new particle spectroscopy, coupling measurements, threshold scans, measurement of spins and other quantum numbers. Accurate jet energy reconstruction will 
  be crucial for this physics programme, driving the need for fine-grained calorimeters and the particle flow approach to calorimetry at CLIC. 

  This paper describes the performance of high granularity particle flow calorimetry at high energies and 
  with the challenging background conditions present at CLIC. These studies build on those 
  first reported in the context of the ILC~\cite{ThomsonNimA}. The results presented here are based on
  full GEANT4 simulations of the CLIC\_ILD~\cite{CLIC_ILD_GeoNote} detector concept considered in the 
  CLIC conceptual design report~\cite{CLIC-CDR}. Results for the 
  CLIC\_SiD~\cite{CLIC_SiD_GeoNote} detector concept were comparable~\cite{PFA_LCD_NOTE}. 

  \section{The Particle Flow Concept}
  Many of the interesting physics processes at CLIC will produce final states containing multiple jets, which may be accompanied by charged leptons and/or missing transverse momentum. In order 
  to perform precision physics measurements, it is vital to be able to reconstruct the invariant masses of the jets; accurate jet mass measurements are a powerful 
  tool for both reconstruction and identification of physics events. The goal for jet energy resolution at CLIC is that it should be sufficient to allow separation of the hadronic decays 
  of $\Wboson$ and $\Zzero$ bosons through the reconstruction of their di-jet invariant masses. This sets a challenging jet energy resolution goal of \mbox{$\sigma_{E}/E\lesssim5-3.5\,\%$} for
  \mbox{$50\,\mathrm{GeV}-1\,\mathrm{TeV}$} jets,
  which is unlikely to be achievable with a traditional approach to calorimetry~\cite{ThomsonNimA}.
  
  Measurements of jet fragmentation at LEP provide detailed information about the particle composition of jets\,\cite{LEPExpt1, LEPExpt2}. In a typical jet, approximately 62\,\% of the energy is carried
  by charged particles (mainly hadrons), whilst 27\,\% is carried by photons, 10\,\% by long-lived neutral hadrons and 1.5\,\% by neutrinos. In a traditional approach to calorimetry, the jet energy
  would be obtained from the energies deposited in the electromagnetic and hadronic calorimeters (ECAL and HCAL respectively). This means that 72\,\% of the energy of a typical jet would be 
  measured with a precision limited by the relatively poor HCAL resolution of \mbox{$\gtrsim 55\%/\sqrt{E/\mathrm{GeV}}$.}
  
  The particle flow approach to calorimetry aims to improve jet energy measurements by reconstructing the four-vectors of all visible particles in an event. 
  The reconstructed jet energy is then the sum of the energies of the individual particles in the jet. 
  At LEP, ALEPH used particle flow techniques~\cite{ALEPH} to improve the energy resolution for hadronic events.
  However, due to the relatively low granularity of the calorimeters, energy depositions from neutral hadrons still had to be identified as significant excesses of calorimetric energy 
  compared to the associated charged particle tracks. Particle flow techniques are also being used by CMS~\cite{CMS} at the LHC.

  The linear collider detector concepts extend the particle flow approach by using 
  fine-granularity calorimeters and sophisticated 
  software algorithms to accurately trace the individual paths of particles through the detector. The energy and momentum for each particle can then be determined from the detector subsystem in which
  the measurements are the most accurate. Charged particle momenta are measured in the inner detector tracker, whilst photon energy measurements are extracted from the
  energy deposited in the ECAL, with typical resolution $<20\%/\sqrt{E/\mathrm{GeV}}$. The HCAL is used to measure only the 10\,\% of jet energy carried by long-lived neutral hadrons. Particle
  flow calorimetry can therefore offer a significant improvement to jet energy measurements, compared to traditional calorimetry.
  For jet energies above about 100\,GeV, the jet energy resolution is limited by the mistakes in the assignment of
  the energies to the different reconstructed particles, termed {\em confusion}, rather than the intrinsic resolution of the 
  calorimeters.
  
  \section{Particle Flow Implementation}\label{sec:PFAImpl}
\begin{figure}[t]
  \begin{centering}
  \includegraphics[width=0.39\paperwidth]{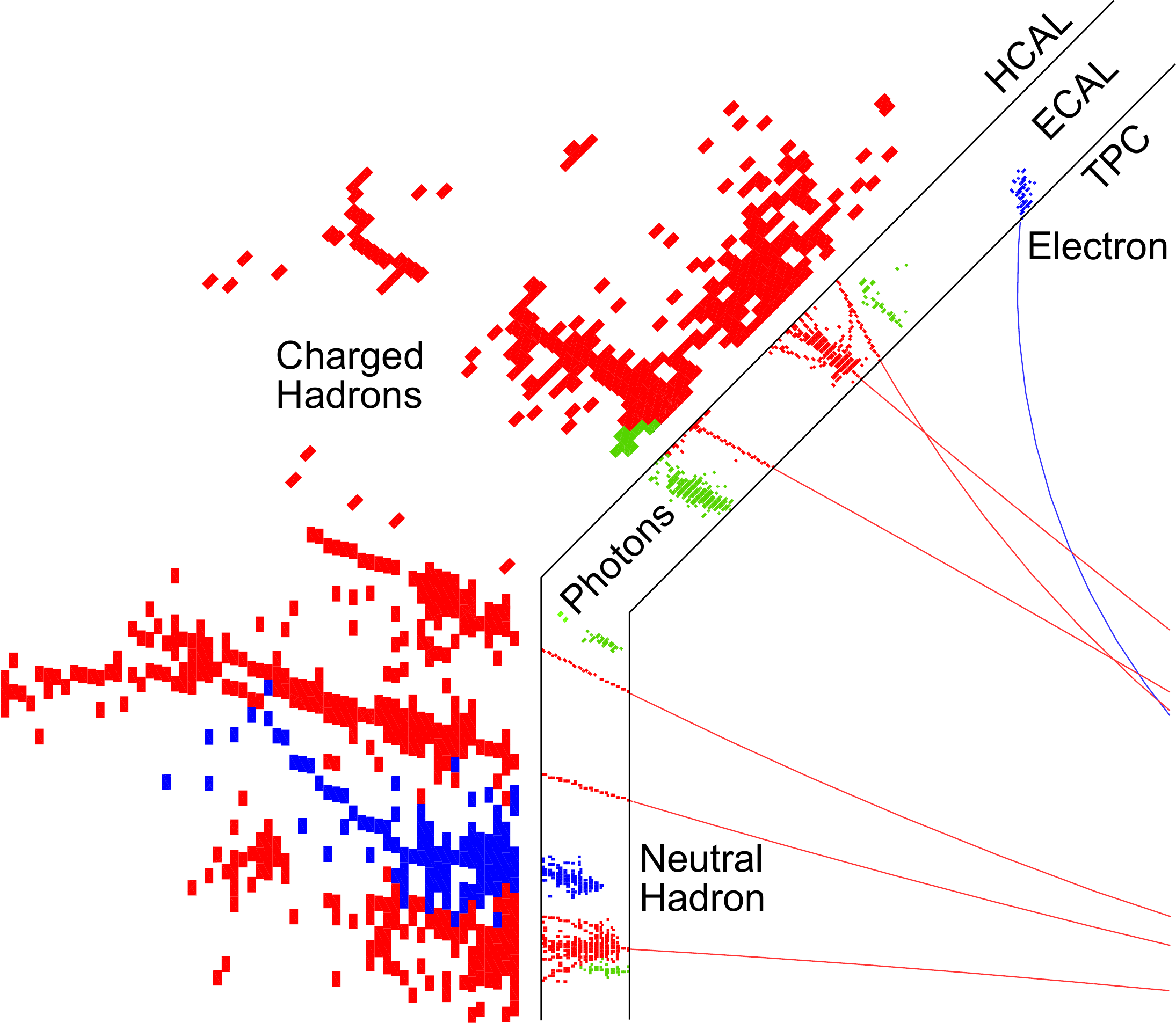}\hspace{2pc}%
  \caption{\label{EventTopology}A typical simulated 250\,GeV jet in CLIC\_ILD, with labels identifying
  constituent particles.}
  \end{centering}
  \end{figure}

  Figure~\ref{EventTopology} shows the typical topology of a simulated 
  250\,GeV jet in the CLIC\_ILD detector concept, with labels identifying a number of the constituent particles. The Figure shows inner detector
  tracks, representing the paths of charged particles in the Time Projection Chamber (TPC). These tracks can be extrapolated by eye and associated with clusters of calorimeter energy
  deposits in the fine granularity ECAL and HCAL. Photons produce energy deposits with characteristic longitudinal and transverse profiles in the ECAL and can be cleanly resolved by eye,
  due to their small transverse spread. HCAL clusters that cannot be associated with TPC tracks represent neutral hadrons. 
  The challenge is to develop software algorithms to automate the reconstruction of the individual particles from the tracks and
  energy deposits in the calorimeters.

  Particle flow calorimetry demands high performance software. The final jet energy resolution is strongly dependent on both the detector granularity and the quality of the particle flow 
  reconstruction algorithms. These algorithms must be able to exploit the granularity to merge together energy deposits from individual particles, with minimal confusion. A logical approach 
  is to implement a series of decoupled pattern-recognition algorithms, each designed to carefully reconstruct a specific particle topology. The implementation of a large number of efficient 
  pattern-recognition algorithms drives the need for a central software framework, which can take care of memory-management and book-keeping issues. Such a framework helps to keep each 
  algorithm simple and focused on its specific pattern-recognition task.
 
  The PandoraPFA C++ Software Development Kit (SDK)~\cite{PandoraPFANew} for particle flow calorimetry is a robust and efficient framework for developing and running algorithms for 
  particle flow reconstruction. It consists of a single framework library and a number of carefully designed Application Programming Interfaces (APIs). It was designed with the twin aims 
  of simplifying the development of efficient pattern-recognition algorithms and allowing easy application of existing algorithms to different detectors, different software environments 
  or even different pattern-recognition tasks. Using the PandoraPFA SDK means that the pattern-recognition reconstruction is divided into three distinct sections, which communicate via 
  the PandoraPFA APIs.

  A PandoraPFA client application uses the APIs to pass details of the tracks and calorimeter cells in an event to the PandoraPFA framework. The framework then creates and manages its own
  lists of self-describing PandoraPFA tracks and cells. These objects are then accessed, in a controlled manner, by the PandoraPFA algorithms. The algorithms control the pattern-recognition 
  reconstruction and determine how the tracks and cells are used to build clusters and, finally, reconstructed particles (termed particle flow objects, or PFOs). Importantly, the algorithms 
  can only access or manipulate the PandoraPFA objects by using APIs to request services from the framework. Typical requests would be to create or delete clusters, merge multiple clusters, 
  split clusters or associate tracks with clusters. This software engineering approach means that the framework can provide all memory-management and book-keeping operations in an efficient 
  and well-tested manner. 
  
  \section{Particle Flow Algorithms}\label{sec:PFAAl}
  The particle flow reconstruction at CLIC uses over 60 different PandoraPFA algorithms and tools. These algorithms are efficient and well-understood. The basic reconstruction operations 
  performed by the default set of PandoraPFA algorithms can be summarised as follows:
  
  \begin{itemize}
  \item Calorimeter cells are clustered using a simple cone-based clustering algorithm. The algorithm works outwards from the innermost cells in the detector, either adding cells to existing 
  clusters or seeding new clusters. Clusters can also be seeded by the projection of inner detector tracks to the front face of the calorimeter.
  \item The clustering algorithm is configured so as to split up the energy depositions from individual particles, rather than risk merging the depositions from multiple particles. The resulting 
  cluster fragments are then carefully merged together by a series of algorithms that implement well-motivated topological rules.
  \item Calorimeter clusters are associated to inner detector tracks, by comparing the properties of the clusters (results of linear fits, helix fits, etc.) to the projected track positions
  and directions at the front face of the calorimeter.
  \item If the energy of a calorimeter cluster does not match the associated track momentum, the clustering can be reconfigured by the statistical reclustering algorithms. These use a 
  series of differently configured clustering algorithms in order to investigate the possible cluster configurations achievable with the relevant cells and tracks. The cluster configuration 
  offering the best track-cluster compatibility is selected.
  \item Fragment-removal algorithms aim to remove neutral clusters that are really fragments of charged clusters. The algorithms search for evidence of association between nearby clusters, whilst 
  also considering the changes in track-cluster compatibility that would occur if the clusters were merged.
  \item Particle flow objects are formed. If a particle is composed of both tracks and clusters, its properties are extracted from measurements of the tracks. For neutral particles, calorimeter
  information is used.
  \item Particle identification algorithms are used throughout the reconstruction.
    \end{itemize}
  
  The original ideas for these algorithms are described in more detail in~\cite{ThomsonNimA}. Since this publication, the algorithms have been re-examined and re-implemented using the PandoraPFA SDK. 
   This new software implementation was used for
  all the physics performance studies presented in the CLIC conceptual design report~\cite{CLIC-CDR}.
  It has been demonstrated~\cite{PandoraPFANew} that the new implementation achieved identical physics performance, whilst offering sizeable reductions in memory usage
  and CPU-time requirements. The re-implementation also offered a number of software engineering improvements, ensuring that the code for each algorithm is simple, efficient
  and easy to maintain. Following the re-implementation of the original algorithms, further development focused on
   improving the reconstruction of high energy jets (above 250\,GeV) in the presence of
  machine-induced background.
  
  For very high energy jets, the energy deposits from particles within a jet inevitably overlap. In this case the pure
  pattern-recognition approach of particle flow reconstruction starts to break down. At this point, more emphasis
  is given to an energy flow approach, based on the consistency of track momenta and associated calorimeter
  cluster energies. Improvements to the statistical reclustering algorithms ensure a smooth transition from 
  pure particle flow to energy flow. These statistical reclustering algorithms can now be 
  cleanly divided into the following categories:
  \begin{itemize}
  \item Algorithms that deal with multiple inner detector tracks associated to a single calorimeter cluster. These algorithms must try to split up the cluster.
  \item Algorithms that deal with clusters with energies significantly greater than the momentum of their associated inner detector tracks. These algorithms must also try to split up the clusters.
  \item Algorithms that deal with inner detector tracks with momentum significantly greater than the energy of their associated clusters. These algorithms must bring in nearby `neutral'
  clusters and reconfigure the division of cells into clusters.
  \end{itemize}

  In order to ensure suitable track-cluster compatibility (an indication that confusion in the reconstruction is minimal), associated tracks and clusters must pass algorithms 
  from each of the above categories without requiring modification. If changes are made, the new track-cluster configuration must also pass the full set of consistency checks. If, and only if, 
  no consistent track-cluster configuration can be found, a `Forced Clustering' algorithm is used to implement a transition to energy flow calorimetry. This algorithm selects calorimeter cells 
  along a helix projection of the inner detector track until a new cluster with energy matching the track momentum is formed. The remnant hits are clustered independently and will typically 
  form neutral particles. This managed transition to energy flow, which occurs only when algorithms cannot find a natural clustering configuration, proves to be important at high jet energies.

  Further improvements for high energy reconstruction include new algorithms to identify MIP-like sections of calorimeter clusters and to track these sections through electromagnetic clusters 
  in the ECAL. Improvements to the photon reconstruction have also been made, placing a series of careful selection cuts on ECAL clusters to ensure that the clusters match the expected longitudinal
  and transverse profiles for electromagnetic showers. Finally, particle identification functionality was added to tag photons and charged leptons. The particle identification information
  is actively used in the pattern-recognition reconstruction, aiding decisions when considering how to process certain clusters or reconstructed particles.

     \section{Background Considerations}
 
The experimental environment at CLIC differs from that at previous 
$\epem$ colliders such as LEP:
\begin{itemize}
\item The high bunch-charge density, related to the small beam size at the
interaction point, means that the electrons and positrons radiate strongly
in the electromagnetic field of the other beam, an effect known as
beamstrahlung.
\item There are significant beam related backgrounds. The pile-up of approximately 3.2
 multiperipheral $\gghadrons$ ``mini-jet'' events per bunch crossing is an important
consideration in particle flow reconstruction at CLIC.
\item The CLIC beam consists of bunch trains of 312 bunches with a train
repetition rate of 50\,Hz. Within a bunch train, the bunches are separated
by 0.5\,ns. The short time between bunches means that a detector will
inevitably integrate over a number of bunch crossings.
\end{itemize} 
If it is assumed that all detector systems integrate over
10\,ns, the $\gghadrons$ background corresponds to 1.2\,TeV of energy deposited in the calorimeter systems.
This large background presents a challenge for particle flow reconstruction at CLIC. However, the combination of
high granularity particle flow reconstruction and timing information can be used to greatly suppress this background.

 \subsection{Background suppression}
 
 The PandoraPFA algorithms aim to reconstruct the four-vectors of all visible particles in an event. A jet-finding algorithm can
 then be used to identify individual jets, with the jet energy reconstructed as the sum of the energies of the 
 reconstructed particles (PFOs) in the jet. 
 However, in  the presence of beam-induced background, it is vital to identify and remove the background particles 
 before the jet-finding stage. Based on full GEANT4 simulation studies, a realistic strategy for mitigating the effects of
 background was developed. This is based on the reconstructed $\pT$ of the individual PFOs and timing information.
 In these studies a 1\,ns timing resolution was assumed for hits in the calorimeters and all calorimeter and tracker 
 hits within a 10\,ns window, starting at the bunch crossing containing the physics event, were included in the reconstruction. 
 Following the reconstruction of inner detector tracks, a \textsc{CLICTrackSelector} event processor was used to remove poor quality and fake
 tracks from events reconstructed in the presence of background. This processor examines the number of track hits in each of the tracking 
 subdetectors and applies simple quality cuts. It also places a cut on the arrival time of the tracked particle at the front face of 
 the calorimeter. Tracks are rejected if the arrival time, calculated using a helix fit to the track, differs by more than 50\,ns from a
 simple straight-line time-of-flight calculation. The \textsc{CLICTrackSelector} is an important tool for background suppression and is used 
 whenever background is added to an event and whenever any background suppression cuts are applied.

 The main benefit of particle flow reconstruction is that it clusters the calorimetric energy in the detector into individual particles, which
 can then be identified as being from background or from the underlying hard interaction.  
 Because hadronic showers take a finite time to develop, it is not possible to simply place very tight timing requirements on the individual
 hits prior to the reconstruction. Instead, the strategy is to perform particle flow reconstruction and then use the properties
 of the PFOs to either reject or retain the PFO. Each calorimeter cluster associated with a reconstructed PFO contains many individual
 hits. The hits are distributed in time due to the assumed time resolution and the development time of the shower. Furthermore, the
 clusters may contain hits from more than one particle; this is particularly important for the forward region, where the background from
 $\gghadrons$ is the greatest. For these reasons, timing cuts are applied at the cluster level using a robust mean of the individual hit
 times. The cluster time is calculated by first determining the median time of all hits in the cluster and then ignoring the times of the outlying 
 10\% of hits. The times of the remaining hits are used to calculate an energy-weighted mean value, which is assigned to the cluster. 
 If a cluster contains sufficient ECAL hits, only ECAL hits are considered in this calculation.
 With a single hit resolution of approximately 1\,ns, 
 mean cluster times with an accuracy much better than 1\,ns are obtained.
  
\begin{figure}[!t]
   \begin{center}
     \includegraphics[width=0.49\textwidth]{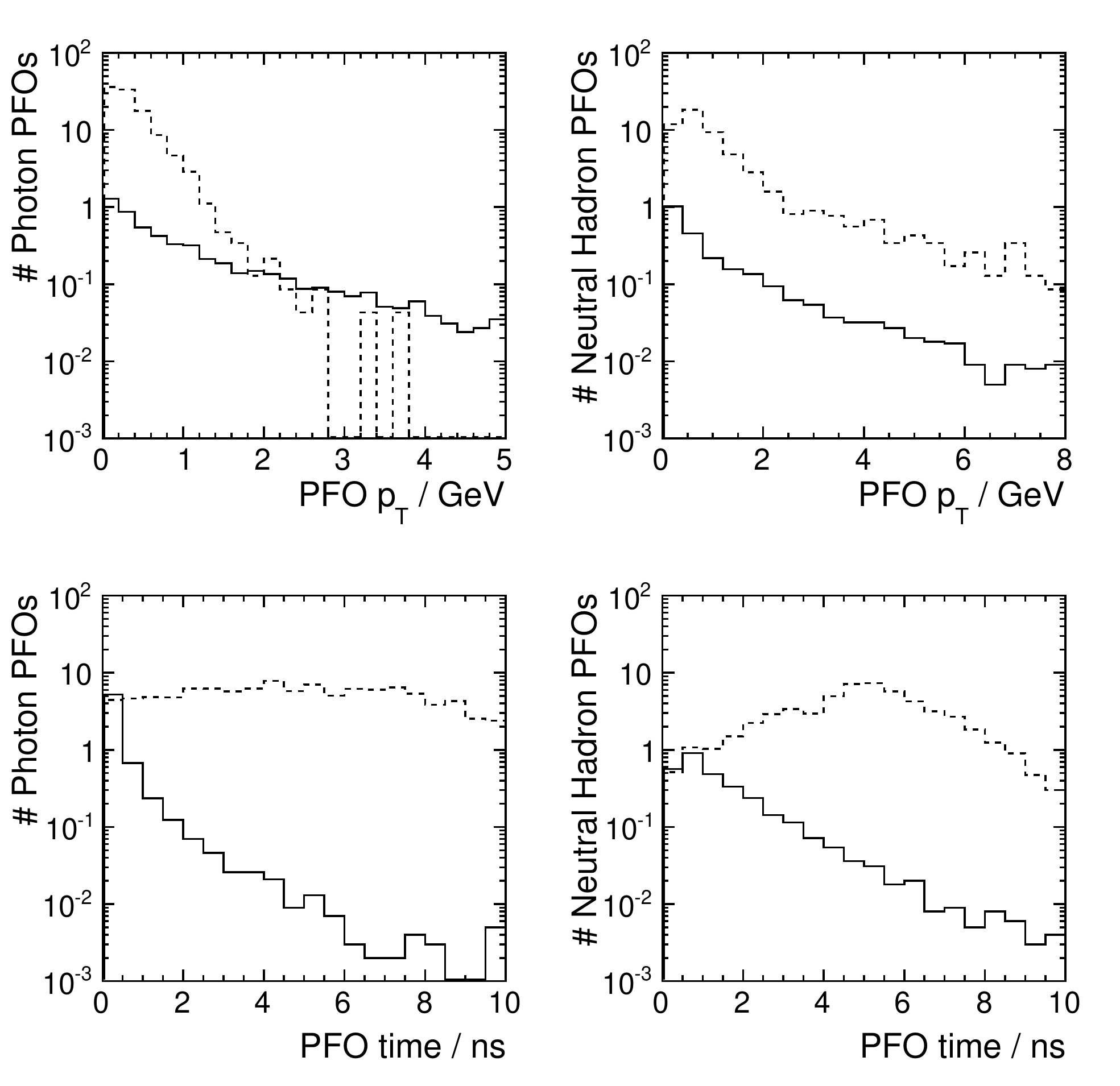}
     \caption{\label{fig:ptE} The PFO time and $\pT$ distributions for reconstructed photon and neutral hadron PFOs.
                          The solid histograms show the distributions for \mbox{$\Zzero\Zzero\rightarrow\neutrino\Aneutrino\Qq\AQq$} events at $\roots=3\,\TeV$ and the
                          dashed histograms show the distribution for pile-up from $\gghadrons$.  }
    \end{center}
  \end{figure}

The PFOs from the physics event have a range of $\pT$ values and have times close to $t_0$, where $t_0$ is the time of
the bunch crossing containing the event; identified in the high-level software trigger. The PFOs from the $\gghadrons$ background
tend to be at lower values of $\pT$ and are distributed in time across the 10\,ns reconstruction window. These two features allow
background PFOs to be separated from signal PFOs. This is only made possible by the highly granular calorimeters. In a more 
conventional calorimeter, the  energy deposits from signal and background would overlap. Figure~\ref{fig:ptE} shows the
$\pT$ and time distributions for photon and neutral hadron PFOs in the endcap regions, for both signal and background. 
The two variables are largely independent. As expected, the time distribution for background photon PFOs is essentially
flat. The reason that the background neutral hadron PFOs tend to peak at 5\,ns is that the reconstructed clusters are
often composed of energy deposits from multiple background particles, which is also reflected in the harder energy spectrum.

The optimal treatment of the background is likely to depend on the physics process being studied. For example, for final state processes which result in forward jets, harder timing
cuts are likely to be beneficial. For this reason sets of background rejection cuts were developed. Cuts were only applied in the regions of $\pT$ and time where there was
a significant background component, $\pT<4.0$\,GeV for photon and charged PFOs and $\pT<8.0\,\GeV$ for neutral hadrons.
Tighter cuts were applied in the far forward region, $|\cos\theta|>0.975$, where background PFOs tend to be formed from energy
deposits from multiple background particles. Tighter timing cuts were applied to PFOs with $\pT<0.75\,$GeV where there is a large contribution
from background.  The initial loose timing cuts,
within the different $\pT$ regions,
were chosen to have minimal impact on PFOs from $500\,\GeV$  jets.  Subsequently the cuts were optimised by 
modifying the timing cuts for each type of PFO in steps of 0.5\,ns.   
The final version of the \textsc{CLICPfoSelector} algorithm has three available configurations, which apply \texttt{Default},  \texttt{Loose}
and \texttt{Tight} cuts to the PFOs. The selection cuts for the \texttt{Loose} and \texttt{Tight} configurations are listed in Table~\ref{tab:PfoSelection}. 
The \texttt{Tight} cuts also include the requirements that $\pT>0.2\,\GeV$ for photon PFOs and $\pT>0.5\,\GeV$ for neutral 
hadron PFOs; these additional cuts were found to be more effective than further tightening the timing cuts.

  \begin{table*}[t]
    \begin{center}
      \caption{Cuts applied by the \textsc{CLICPfoSelector} in the \texttt{Loose} and \texttt{Tight} configuration modes.
	\label{tab:PfoSelection}}
      \begin{tabular}{ c c c c c }
	\toprule
	& \multicolumn{2}{c}{ \texttt{Loose} configuration}  & \multicolumn{2}{c}{\texttt{Tight} configuration} \\
	\midrule
	\bf{Region} & \bf{\emph{p}}$\bf{_{T}}$ \bf{range} [GeV] & \bf{Time} [ns] & \bf{\emph{p}}$\bf{_{T}}$ \bf{range} [GeV] & \bf{Time} [ns]\\
	\midrule
	\multicolumn{5}{c}{\bf{Photons}} \\
	\midrule
	Central                   & $0.75 \leq \pT < 4.0$  & $t < 2.0$ & $1.0 \leq \pT < 4.0$   & $t < 2.0$\\
	$|\cos(\theta)|\leq0.975$ & $0 \leq \pT < 0.75$    & $t < 2.0$ & $0.2 \leq \pT < 1.0$   & $t < 1.0$\\
	Forward                   & $0.75 \leq \pT < 4.0$  & $t < 2.0$ & $1.0 \leq \pT < 4.0$   & $t < 2.0$\\
	$|\cos(\theta)|>0.975$    & $0 \leq \pT < 0.75$    & $t < 1.0$ & $0.2 \leq \pT < 1.0$   & $t < 1.0$\\
	\midrule
	\multicolumn{5}{c}{\bf{Neutral hadrons}} \\
	\midrule
	Central                   & $0.75 \leq \pT < 8.0$  & $t < 2.5$ & $1.0 \leq \pT < 8.0$   & $t < 2.5$\\
	$|\cos(\theta)|\leq0.975$ & $0 \leq \pT < 0.75$    & $t < 1.5$ & $0.5 \leq \pT < 1.0$   & $t < 1.5$\\
	Forward                   & $0.75 \leq \pT < 8.0$  & $t < 2.5$ & $1.0 \leq \pT < 8.0$   & $t < 1.5$\\
	$|\cos(\theta)|>0.975$    & $0 \leq \pT < 0.75$    & $t < 1.5$ & $0.5 \leq \pT < 1.0$   & $t < 1.0$\\
	\midrule
	\multicolumn{5}{c}{\bf{Charged particles}} \\
	\midrule
	All                       & $0.75 \leq \pT < 4.0$  & $t < 3.0$ & $1.0 \leq \pT < 4.0$   & $t < 2.0$\\
        & $0 \leq \pT < 0.75$    & $t < 1.5$ & $0 \leq \pT < 1.0$     & $t < 1.0$\\
	\bottomrule  
      \end{tabular}
    \end{center}
  \end{table*}

  \begin{figure*}[!htb]
    \begin{center}
      \includegraphics[width=0.49\textwidth]{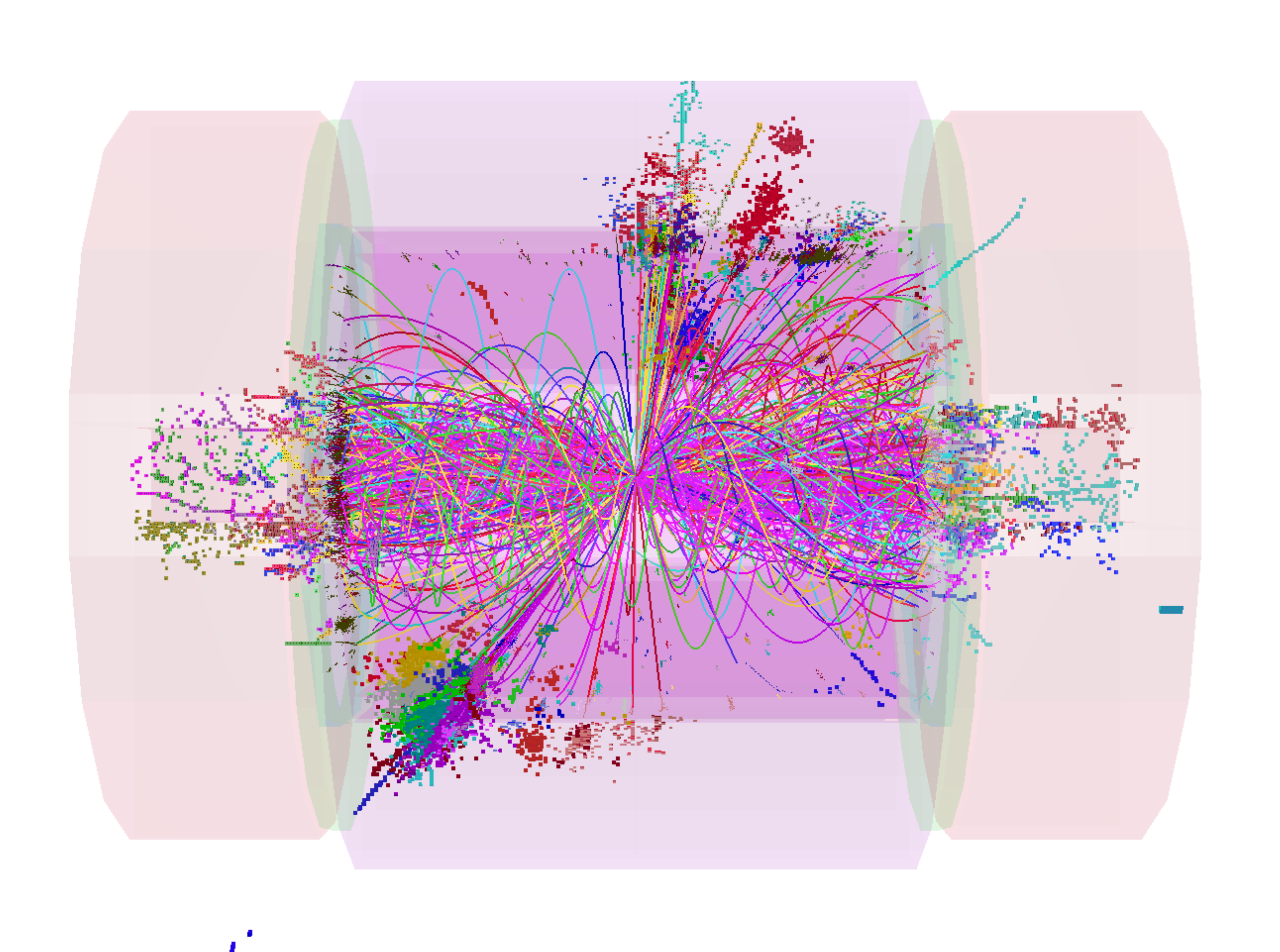}
      \includegraphics[width=0.49\textwidth]{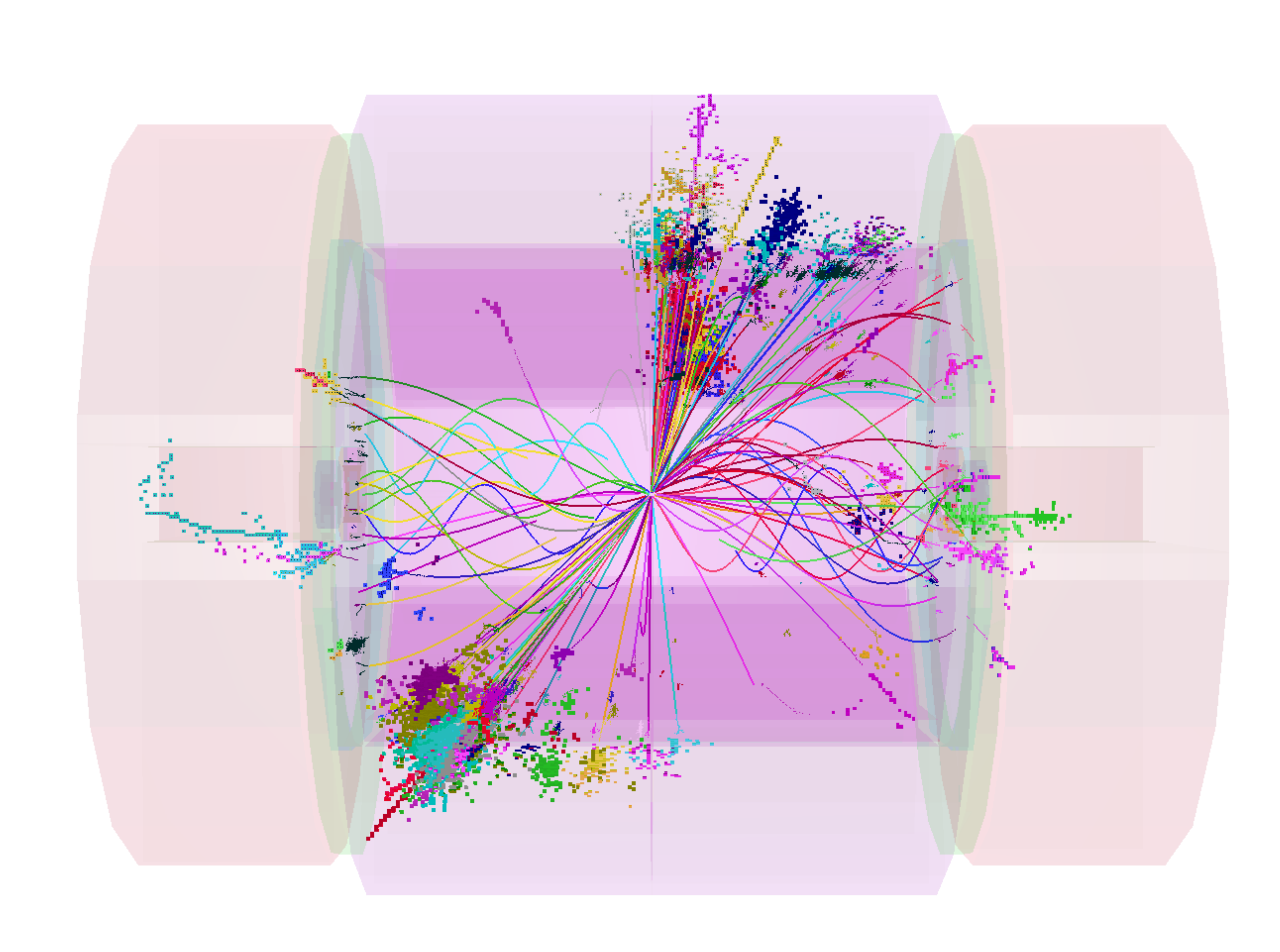}
      \caption{Reconstructed particles for a time window of 10 ns (100 ns in HCAL barrel)
	in a simulated \mbox{$\epem\rightarrow \Higgs^{+}\Higgs^{-} \rightarrow \Qt\AQb\Qb\AQt$}
	event in the CLIC\_ILD detector, with 60 BX of $\gghadrons$ background overlaid (left). The effect of applying \texttt{Tight} 
	PFO selection cuts to the reconstructed particles is shown on the right; the energy deposited in the detector by the background is 
	reduced from 1.2~TeV to the level of 100~GeV.\label{figPFOSelectorEvDisplay}}
    \end{center}
  \end{figure*}

  To illustrate the power of suppressing machine-induced background with the \textsc{CLICPfoSelector}, Figure~\ref{figPFOSelectorEvDisplay} 
  shows the reconstructed particle flow objects for a simulated \mbox{$\epem\rightarrow \Higgs^{+}\Higgs^{-} \rightarrow \Qt\AQb\Qb\AQt$} event at \mbox{$\sqrt{s}$ = 3\,TeV.}
  Assuming a time window of 10\,ns for the silicon detectors, the ECAL and the HCAL endcap and 100 ns for the HCAL barrel, the background from $\gghadrons$ produces an average energy 
  of approximately 1.2\,TeV per event, mostly in the form of relatively low $\pT$ particles at relatively low angles to the beam axis. As a result of the cluster-based \texttt{Tight} 
  timing cuts, the average background level can be reduced to approximately 100\,GeV with negligible impact on the underlying hard interaction. Table~\ref{tab_timing_Eres} summarises the 
  impact of the different PFO selection cuts on the signal and background. The equivalent cut on $\pT$ alone has a much larger impact on the signal than the combination of $\pT$ and timing cuts.

  \begin{table}[!h]
    \centering
    \caption{The impact of the different PFO selections on the reconstructed energy of $\gghadrons$, and of a hadronically decaying $\Wboson$ boson of 500\,GeV energy. The impact of a simple 
             $\pT$ cut is shown for comparison.}
    \begin{tabular}{l r r r}
      \toprule
      Cut & \multicolumn{1}{c}{$\gghadrons$} &\multicolumn{2}{c}{500\,GeV di-jet} \\
      & Energy & Energy & Energy \\
      & (GeV) & (GeV) & loss \\
      \midrule
      \texttt{No cut} & 1210 & 500.2 & 0\% \\
      \texttt{Loose} & 235 & 498.8 & 0.3\% \\
      \texttt{Default} & 175 & 498.0 & 0.5\% \\
      \texttt{Tight} & 85 & 496.1 & 0.8\% \\
      \midrule
      $\pT > \unit[3.0]{GeV}$ & 160 & 454.2 & 9.2\% \\
      \bottomrule
    \end{tabular}\label{tab_timing_Eres}
  \end{table}
  
  \section{A Study of Particle Flow Performance}
  \label{sec:TechnicalStudy}

  A study has been carried out to assess the performance of particle flow calorimetry at CLIC, and to investigate the variation of the performance with the jet energy, the angle and the
  PFO selection cuts. For this purpose, $\Zzero^\prime$ particles were generated at energies ranging from 91\,GeV to 3\,TeV. These are off-shell $\Zzero$ bosons, produced at rest at different
  centre-of-mass energies, which decay into light quarks and typically provide two back-to-back mono-energetic jets. For this initial study, no jet reconstruction was performed and backgrounds 
  were not included; instead, to avoid bias, the full energy deposited in the detector, $E_{jj}$, was analysed. The performance of fully reconstructed physics observables in the presence of 
  background is presented in Section~\ref{sec:PhysicsPerformance}.

  The performance of the particle flow reconstruction was evaluated by calculating the resolution of the jet energy, $E_{j}$. This is defined as:
  \begin{equation}
    \frac{\mathrm{RMS}_{90}(E_j)}{\mathrm{mean}_{90}(E_{j})} = \frac{\mathrm{RMS}_{90}(E_{jj})}{\mathrm{mean}_{90}(E_{jj})} \sqrt{2}
  \end{equation}
  where the RMS$_{90}(E_{jj})$ and the mean$_{90}(E_{jj})$ are calculated from the total reconstructed energy distribution. The RMS$_{90}$ is defined as the RMS in the smallest region of reconstructed 
  energy that contains 90\,\% of the events. It is introduced in order to reduce sensitivity to tails in a well defined manner, because the effects of confusion mean that the PFO energy distribution
  will be inherently non-Gaussian~\cite{ThomsonNimA}.

  \subsection{Jet Energy Resolution}
\begin{figure*}[!htb]
    \begin{center}
      \includegraphics[width=0.49\textwidth]{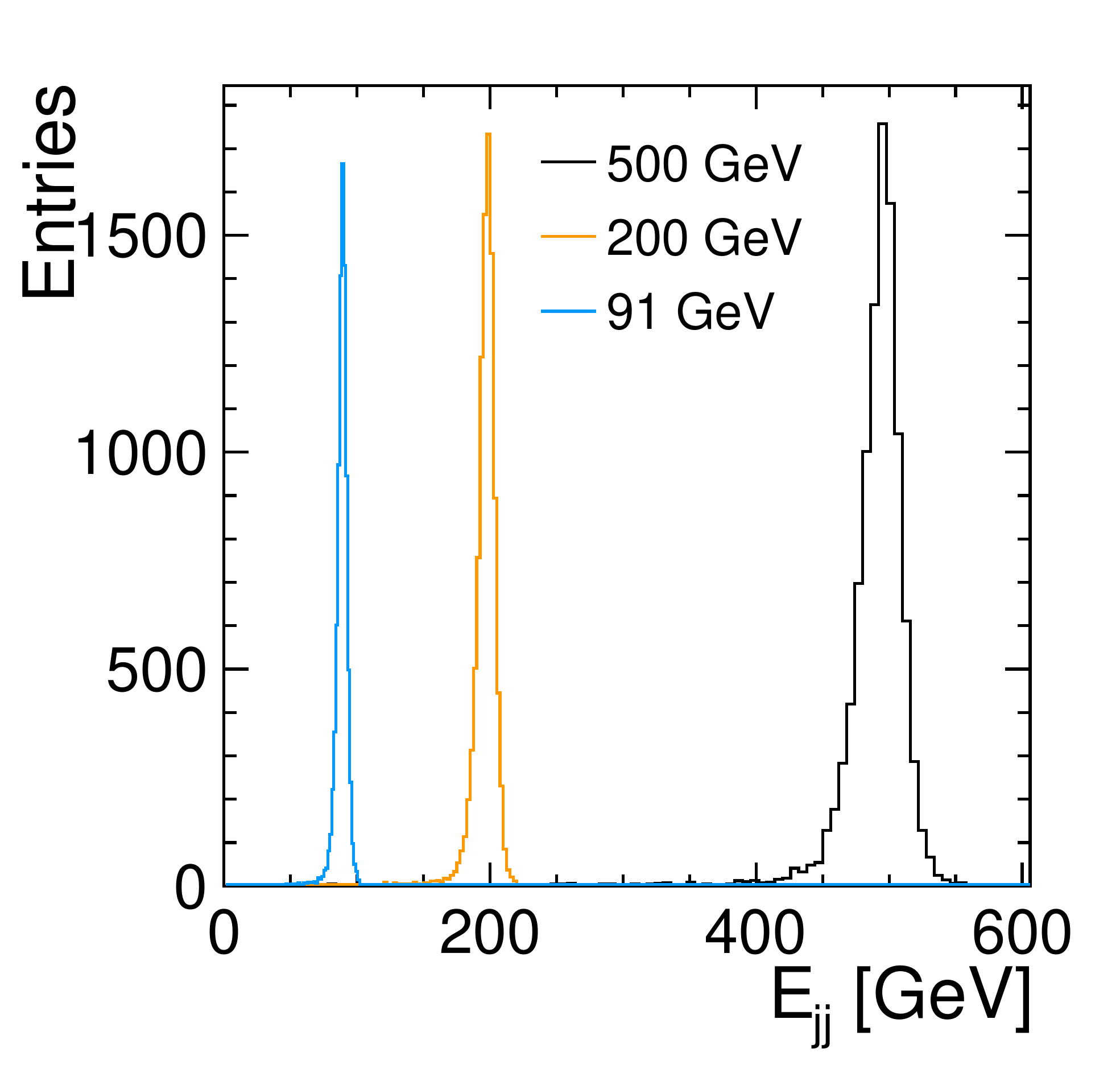}
      \includegraphics[width=0.49\textwidth]{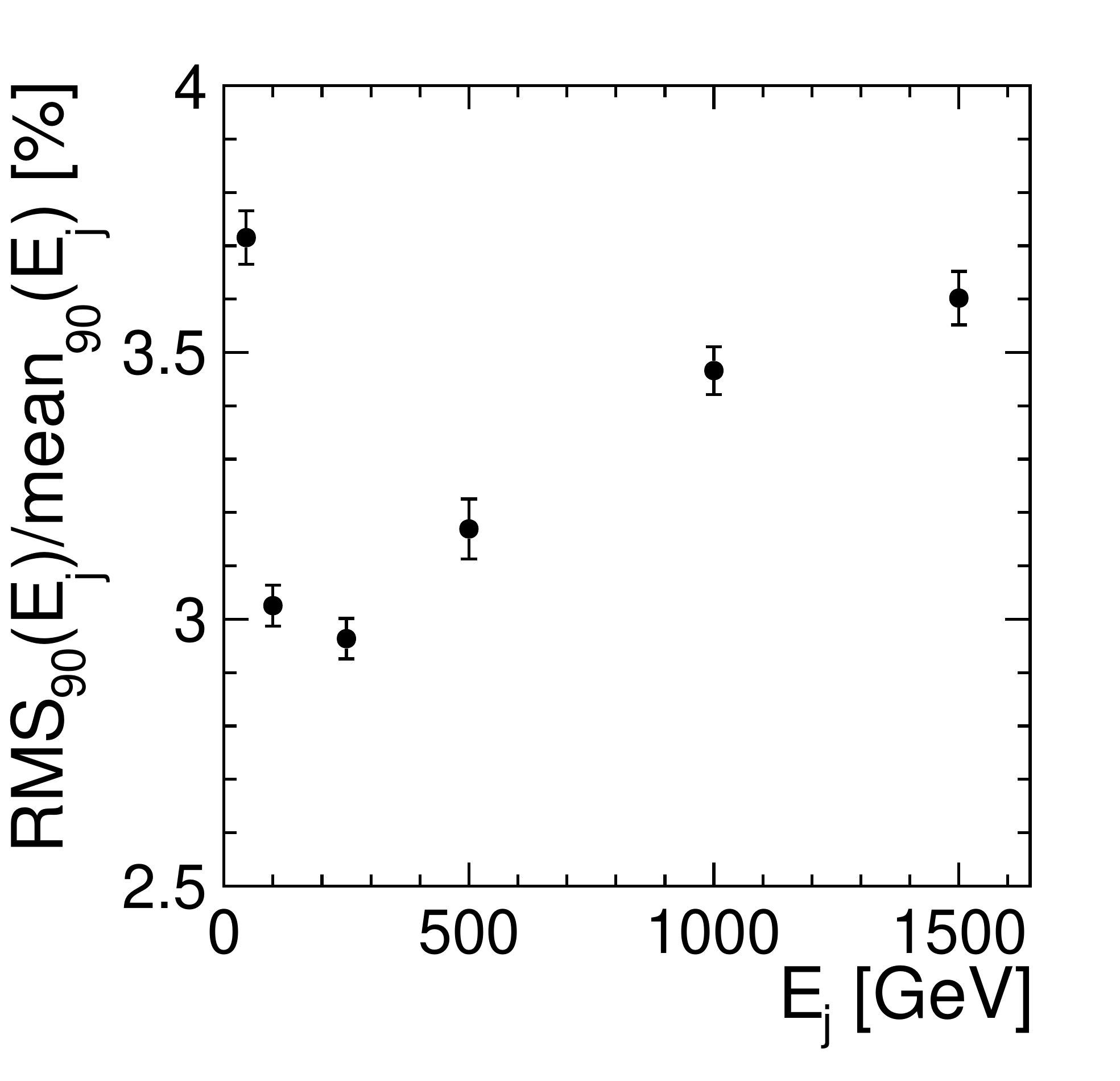}
      \caption{The total reconstructed energy for three centre-of-mass energies (left) and the jet energy resolution as function of the jet energy (right) for jets from 
               simulated \mbox{$\Zzero^\prime\rightarrow\text{q}\overline{\text{q}}$} decays. The jet energy resolutions represent only jets in the barrel region of the detector. \label{figZUDS_Eres_vsE}}
    \end{center}
  \end{figure*}

  As the jet energy increases, the jets become narrower and it becomes more difficult to distinguish individual particles. Particle flow calorimetry can turn into energy flow calorimetry and the 
  effects of confusion will dominate the jet energy resolution. Figure~\ref{figZUDS_Eres_vsE} shows the distribution of the reconstructed jet energy for a number of different jet energies. The 
  Figure also shows the variation of the jet energy resolution as a function of the jet energy for jets in the barrel region of the detector, defined by $|\cos(\theta)|<0.7$, where $\theta$ is the 
  polar angle of the generated quarks. Compared to previous results obtained in~\cite{ThomsonNimA}, significant improvements to the jet energy resolution have been achieved with the new PandoraPFA 
  algorithms. The improvements are particularly noticeable at high jet energies; the jet energy resolution is better than 3.7\,\% over the range of jet energies considered. Whilst the results 
  presented here are for the barrel region of the detector, a jet energy resolution of 4\,\% or better is achieved in the endcap region, only worsening slightly at $0.95< |\cos{\theta}| <0.975$, 
  corresponding to angles as small as 12.8$^{\circ}$. The full variation of the jet energy resolution with polar angle is shown for both CLIC detector concepts in~\cite{CLIC-CDR}.

%

    \subsection{PFO Selection Cuts}
\begin{figure*}[!htb]
    \begin{center}
      \includegraphics[width=0.49\textwidth]{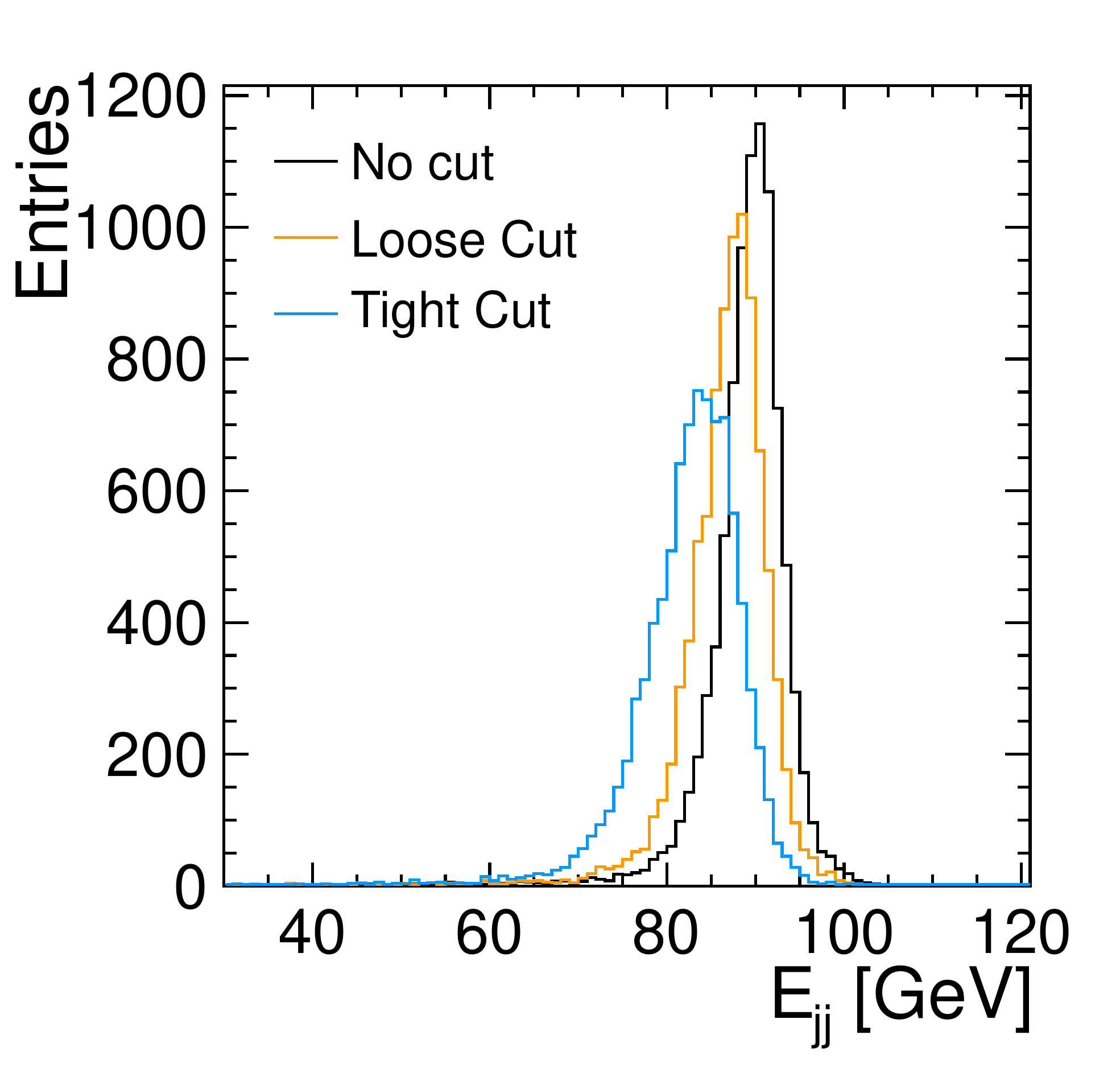}
      \includegraphics[width=0.49\textwidth]{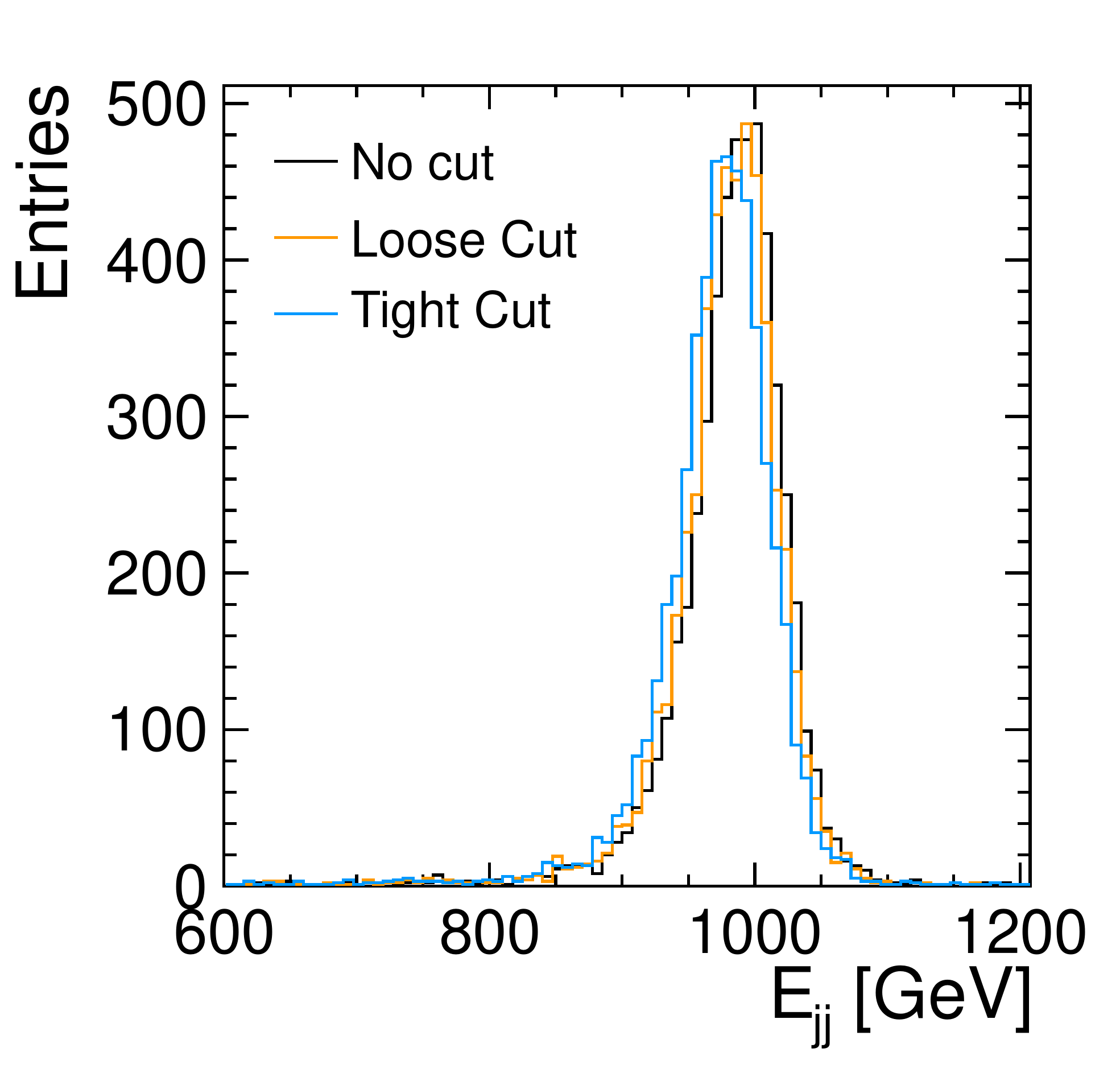}
      \caption{The impact of the the PFO selection cuts on the total reconstructed energy at $\sqrt{s}=91$\,GeV (left) and $\sqrt{s}=1$\,TeV (right), for jets from 
               simulated \mbox{$\Zzero^\prime\rightarrow\text{q}\overline{\text{q}}$} decays. \label{figZUDS_timing}}
    \end{center}
  \end{figure*}

  In order to assess the impact of the \textsc{CLICPfoSelector} on underlying physics events, the \mbox{$\Zzero^\prime\rightarrow\text{q}\overline{\text{q}}$} events described in Section~\ref{sec:TechnicalStudy} 
  were used again. Without any 
  backgrounds, the $\Zzero^\prime$ events were examined after the application of the different PFO selection cuts. Figure~\ref{figZUDS_timing} shows the sum of the reconstructed PFO energies for 91\,GeV 
  $\Zzero^\prime$ events, before application of PFO selection cuts. Separate distributions are shown after application of the \texttt{Loose} and \texttt{Tight} cuts described in Table~\ref{tab:PfoSelection}. The more 
  stringent the cuts, the more energy is cut away. Equivalent distributions are also shown for 1\,TeV $\Zzero^\prime\rightarrow\text{q}\overline{\text{q}}$ events, indicating that any negative impact of the PFO 
  selection cuts decreases with energy.

\begin{figure}[!h]
    \begin{center}
      \includegraphics[width=0.49\textwidth]{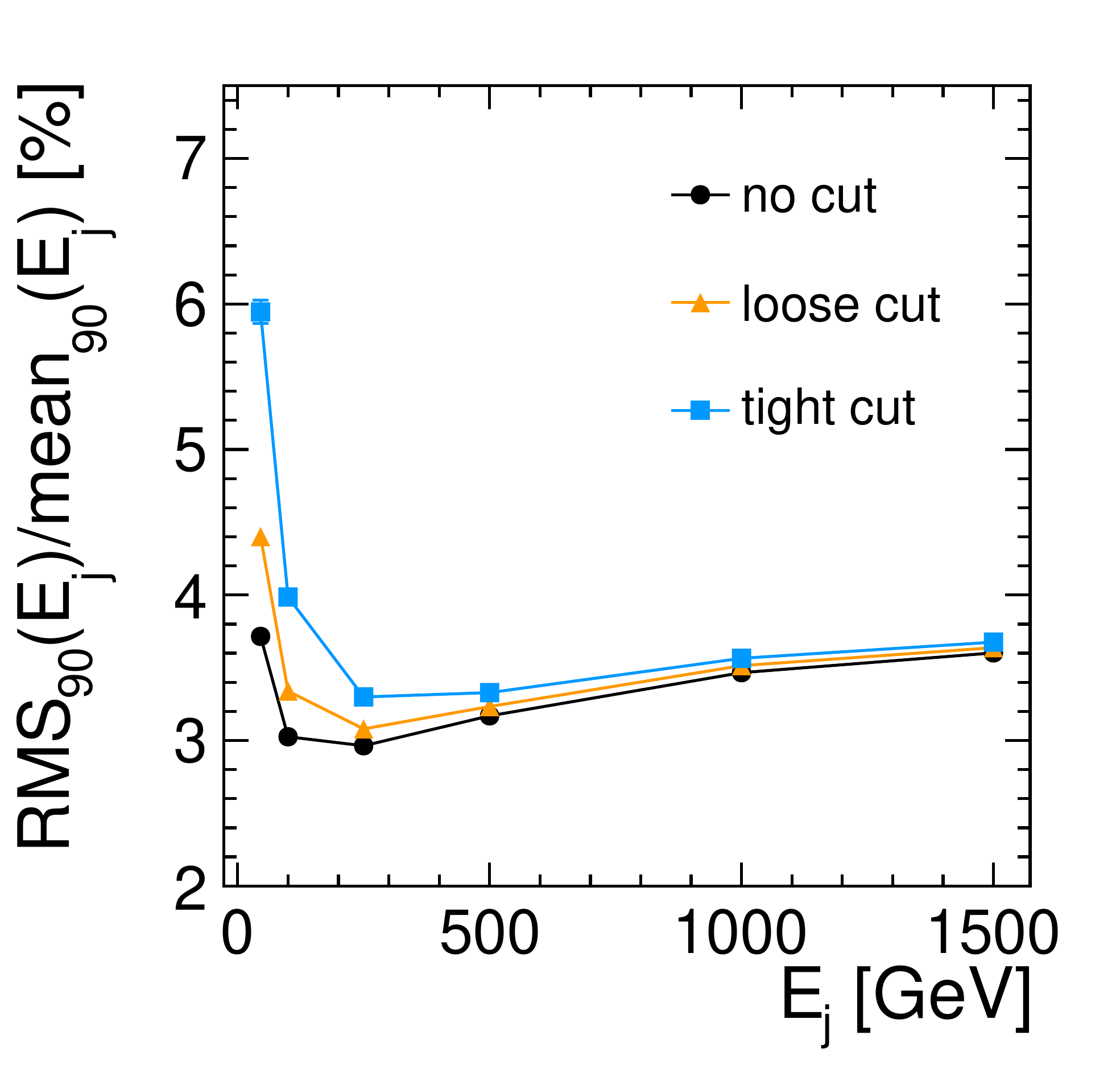}
      \caption{The impact of PFO selection cuts on jet energy resolution for jets from simulated \mbox{$\Zzero^\prime\rightarrow\text{q}\overline{\text{q}}$} decays. \label{figZUDS_timing_Eres}}
    \end{center}
  \end{figure}

  Figure~\ref{figZUDS_timing_Eres} illustrates the impact of the \textsc{CLICPfoSelector} on the jet energy resolution as a function of the jet energy. At low jet energies, the PFO selection cuts have
  a significant impact on the jet energy resolution. As the jet energy increases, the jet energy reconstruction performance becomes the same with or without application of the PFO selection cuts.
   
  \section{Particle Flow Physics Performance}
  \label{sec:PhysicsPerformance}

  \subsection{$\Wboson$ and $\Zzero$ Reconstruction}
  \label{sec:WW}
  In order to study the physics performance achievable using particle flow calorimetry at CLIC, the ability to distinguish between $\Wboson$ and $\Zzero$ particles was examined. The $\Wboson$ 
  events used for this study each contain two $\Wboson$ bosons, one of which decays into a muon and neutrino, whilst the other decays into quarks. Events were generated for $\Wboson$ energies of 
  125, 250, 500 and 1000\,GeV.
  The main background for physics at CLIC is from the $\gghadrons$ mini-jet events. For each bunch crossing (BX), a number of $\gghadrons$ events was superimposed on the physics event; the number of events
  to be overlaid in this way was drawn from a Poisson distribution, assuming a mean of 3.2 events per BX~\cite{BKGR_noteILD}. For each $\Wboson$ energy, samples were produced without background (0~BX)
  and with background overlaid to represent sixty bunch crossings (60~BX). To further study the impact of background on physics performance, additional samples were produced with a safety factor of two 
  in the background estimation, corresponding to a mean of 6.4 $\gghadrons$ per BX.

  All samples were fully simulated and reconstructed. Additional reconstruction and event selection procedures were then applied as described below:
  
  \begin{itemize}
  \item Lepton Removal:\\ 
    The muon and every particle around it within a cone of $|\cos{\alpha}| > 0.9$ was removed, leaving only the hadronic decay of one $\Wboson$ in the event.
  \item Removal of Neutral Fragments:\\
    If background had been overlaid, low energy neutral fragments in the forward region ($|\cos{\theta}| > 0.9$) were carefully removed. Fake neutral PFOs are likely to be created in this region due to
    unsuccessful matches between low $\pT$ curling tracks and associated calorimeter clusters. A minimum energy threshold was applied for neutral PFOs in the region, with a cut ranging between 1 and 8\,GeV,
    depending on the centre-of-mass energy. 
  \item Jet Reconstruction:\\
    Following evaluation of a number of jet reconstruction algorithms, the FastJet library~\cite{FastJet} was selected and the \texttt{kt} algorithm used in exclusive mode to force the event into two jets.
    A scan of the $R$ parameter for the jet cone size was performed at each energy, with and without background, and for each set of \textsc{CLICPfoSelector} cuts: \texttt{Default}, \texttt{Loose} and \texttt{Tight}.
    No significant differences were observed for the different PFO selections. The best $R$ values were found to be 0.7 with background and 1.5 without background.

    The performance of particle flow calorimetry is closely linked to the quality of the jet reconstruction. Use of a single jet reconstruction algorithm for all configurations is not optimal, but does simplify
    the comparison of events with and without background. For this reason, the same jet reconstruction was chosen for use with all samples.
  \item Event Selection:\\
    Both jets were required to have $|\cos{\theta}| < 0.9$, to stay well within the detector acceptance.  In addition, a lower limit on the angle between the two jets was applied, to reject events where the jet
    reconstruction failed. The lower limit depended on the energy of the $\Wboson$ and was based on the expected range of angles for the generated events.
  \end{itemize}

  \begin{figure*}[!htb]
    \begin{center}
      \includegraphics[width=0.49\textwidth]{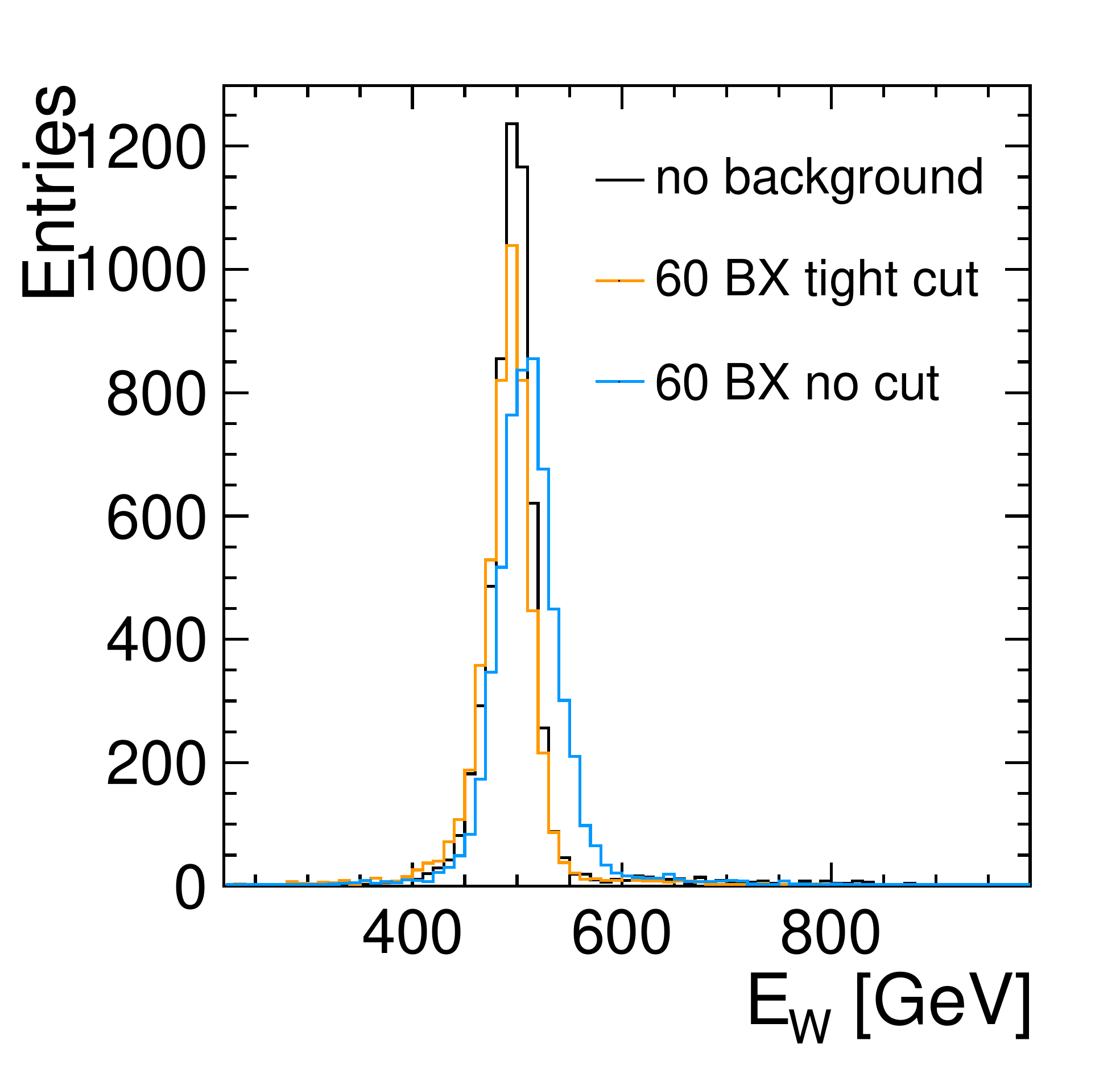}
      \includegraphics[width=0.49\textwidth]{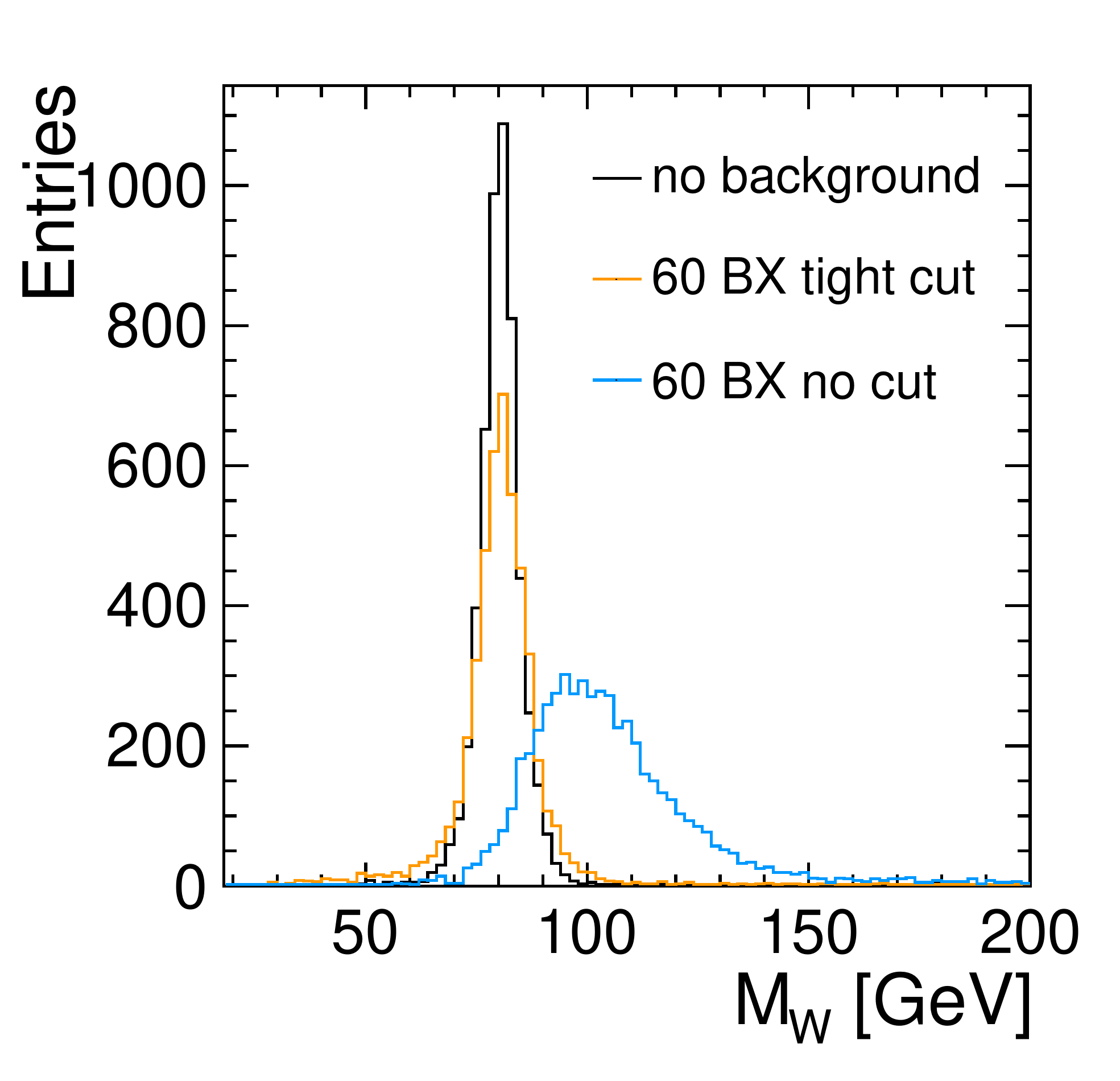}
      \caption{Energy (left) and mass (right) distributions of reconstructed $\Wboson$ at an energy of 500\,GeV. Separate distributions are shown for events without background and for events with 60~BX of 
               background before and after application of \texttt{Tight} PFO selection cuts.
	\label{figWW_PFOSel}}
    \end{center}
  \end{figure*}

  The need for the PFO selection cuts in the presence of background is clearly illustrated in Figure~\ref{figWW_PFOSel},  
  which shows the reconstructed energy and mass distributions for the 500\,GeV $\Wboson$
  samples. Separate distributions are shown for the samples without background, and the 60~BX sample 
  with and without PFO selection cuts. Without the application
  of the PFO selection cuts, many background particles remain in the event and are reconstructed as part of the jet. These particles   
  shift the jet energy distribution to higher values and tend to bias the reconstructed jet axis towards the beam axis. 
  Consequently, the reconstructed di-jet mass distribution is badly distorted. With the PFO selection, a narrow
  $\Wboson$ mass peak is recovered.
  It is found that the differences between the three PFO selection cuts are small and henceforth only the \texttt{Tight} cuts are considered. 

  The energy and
  mass resolutions were studied by calculating the RMS$_{90}$ and mean$_{90}$ for the distributions of jet energy and jet mass. With the overlay of 60~BX of background, the energy and mass distributions 
  were distorted and the tails became more prominent. The RMS$_{90}$ method to calculate the resolution is robust against these changes,
  whilst considering all features of the distribution,
  not just the main part of the peak. The resolutions given in this section are therefore purely a measure of peak quality, but cannot be used directly to assess the power to distinguish between $\Wboson$ and
  $\Zzero$ particles. This is addressed in Section~\ref{sec:sepWZ}.

\begin{figure*}[!htb]
    \begin{center}
      \includegraphics[width=0.49\textwidth]{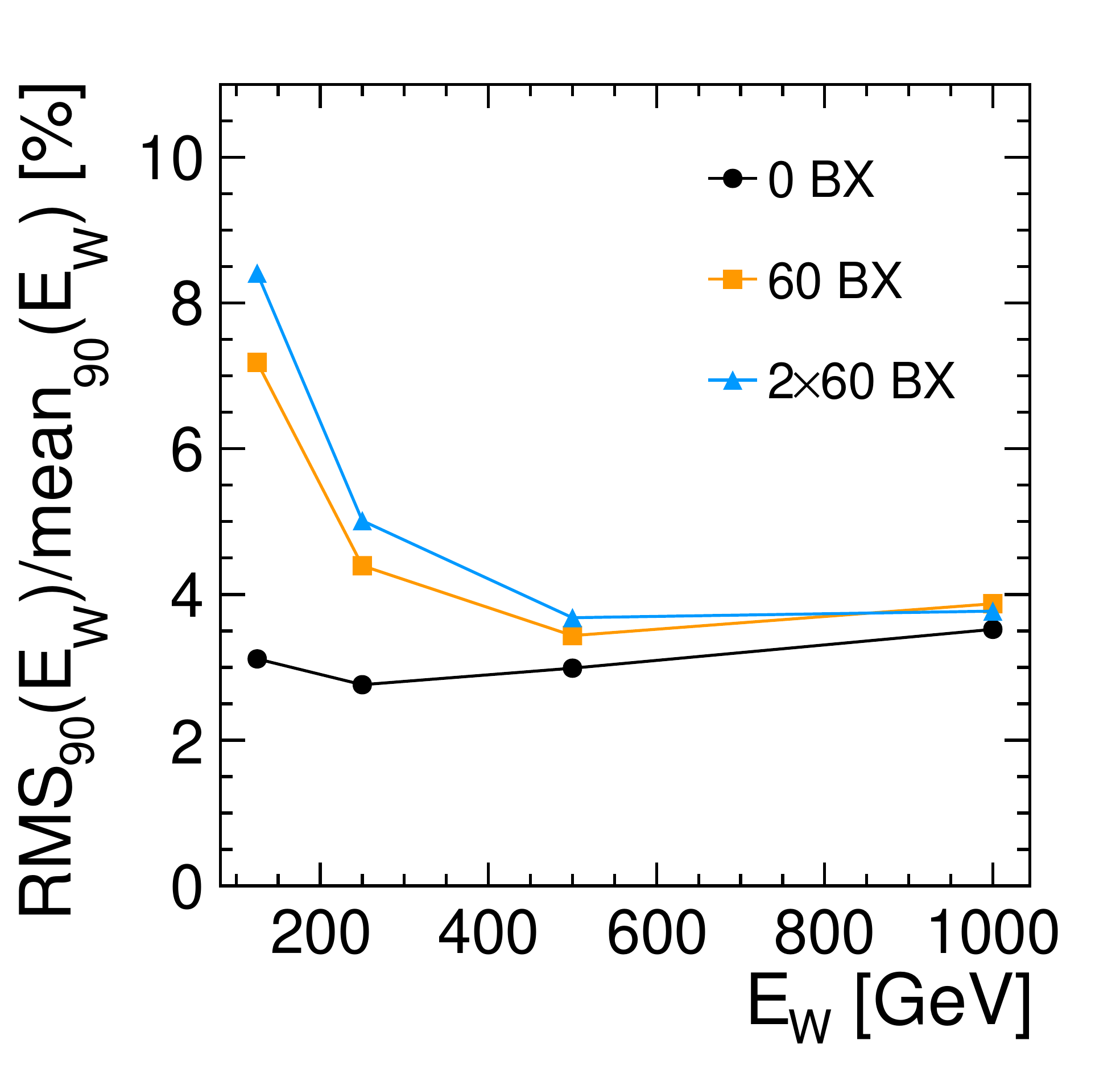}
      \includegraphics[width=0.49\textwidth]{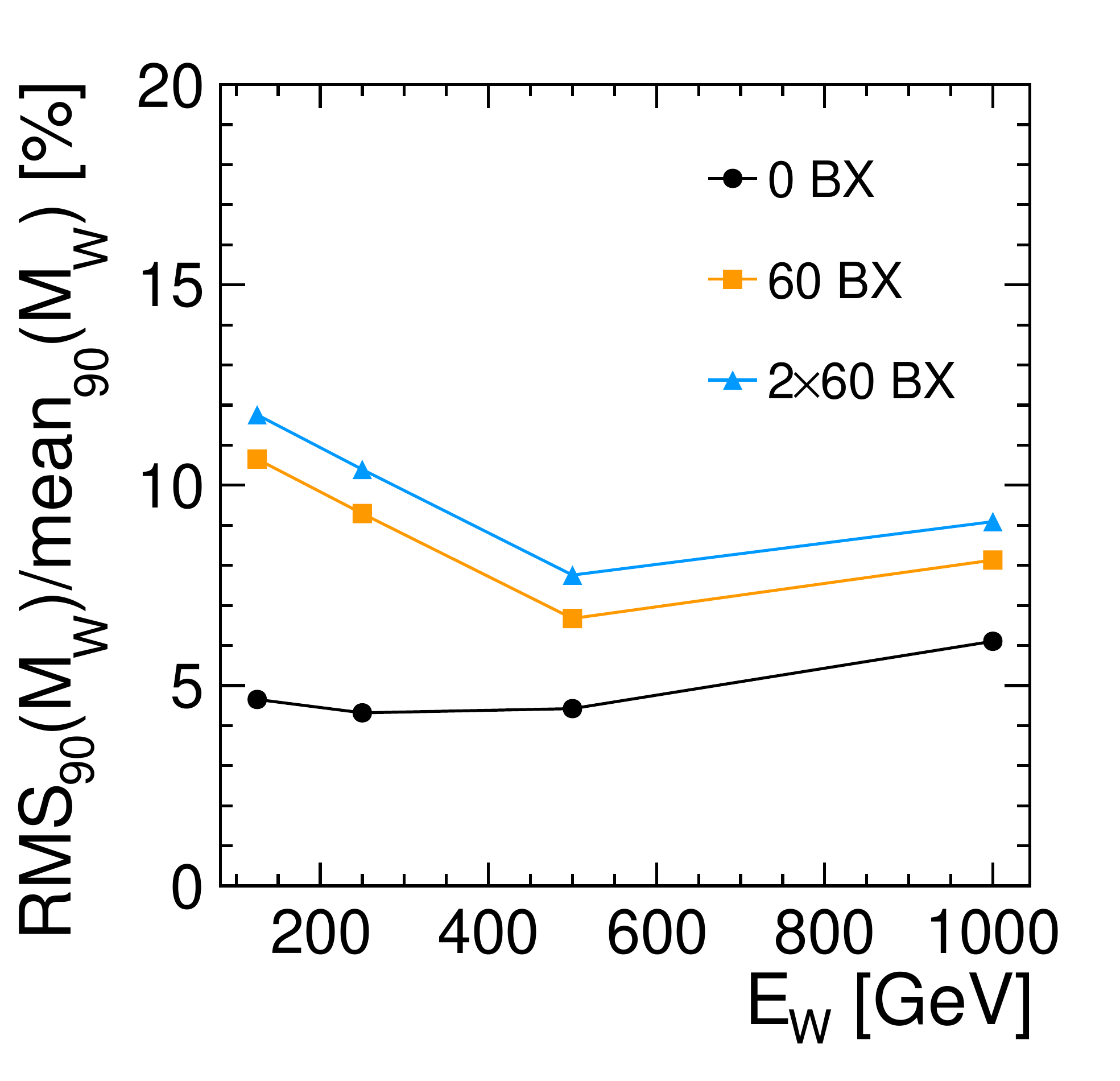}
      \caption{Energy (left) and mass (right) resolutions of reconstructed $\Wboson$, for events without background and for events with 60~BX and 2$\times$60~BX of background.
	\texttt{Tight} PFO selection cuts are used for events with background.\label{figWW_res}}
    \end{center}
  \end{figure*}

  Figure~\ref{figWW_res} shows the energy and mass resolution of the reconstructed $\Wboson$ as a function of the $\Wboson$ energy. Separate distributions are shown for the samples without background, for the 
  samples with 60~BX of background and for the samples with 2$\times$60~BX of background. Without background, the resolutions are comparable to those obtained in the study described in Section~\ref{sec:TechnicalStudy}, 
  without jet reconstruction.
  In the presence of background, the degradation of the energy resolution at lower energies is significant. With increasing $\Wboson$ energy, the impact of the background on the energy resolution becomes rather
  small. As the mass resolution is more sensitive to the jet quality, the background still leads to appreciable degradation of the mass resolution, even at higher energies. The additional degradation upon moving
  from 60~BX to 2$\times$60~BX of background is rather small.

  \subsection{$\Wboson$ and $\Zzero$ Separation}
  \label{sec:sepWZ}
\begin{figure*}[!htb]
    \begin{center}
      \includegraphics[width=0.49\textwidth]{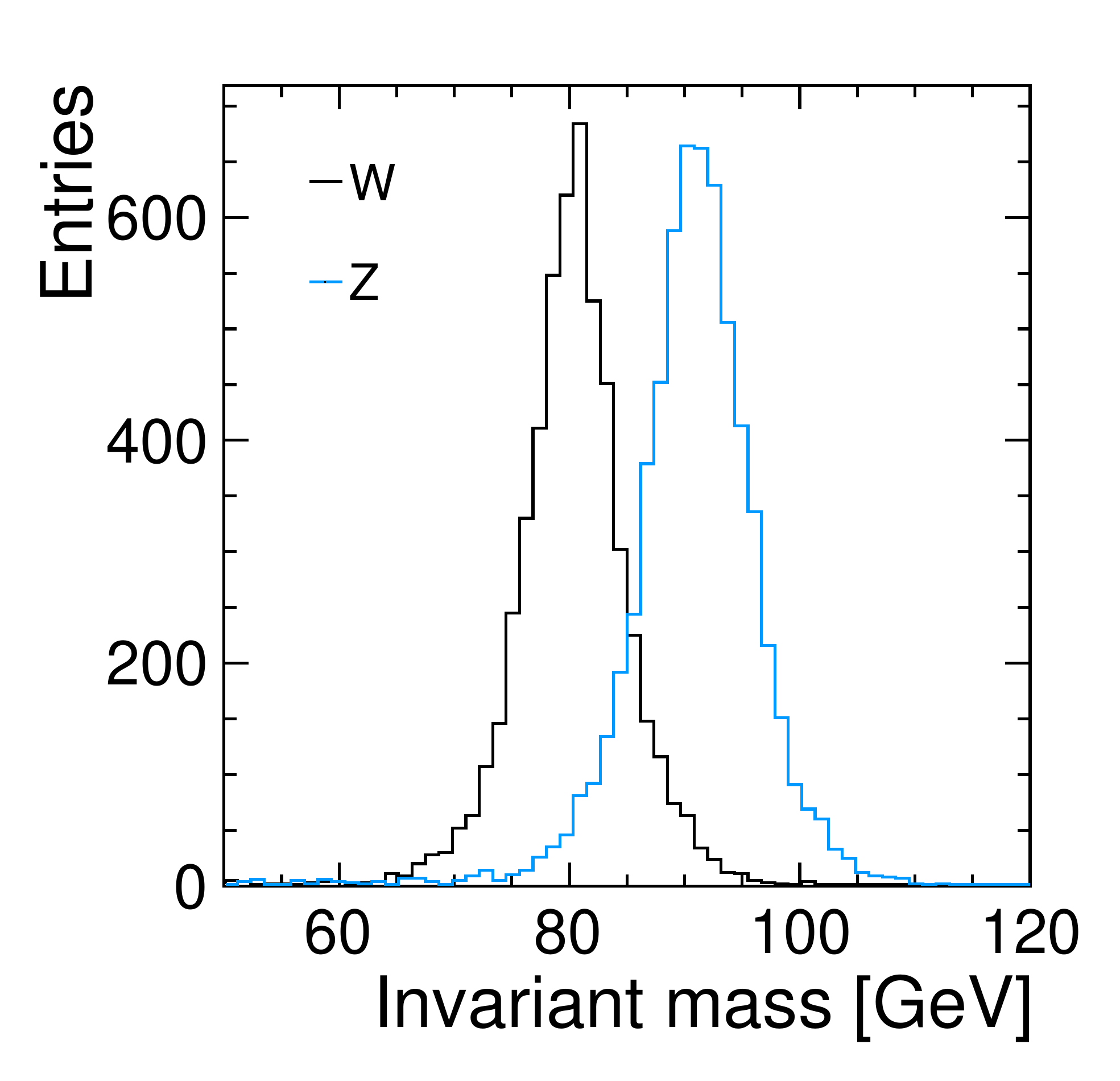}
      \includegraphics[width=0.49\textwidth]{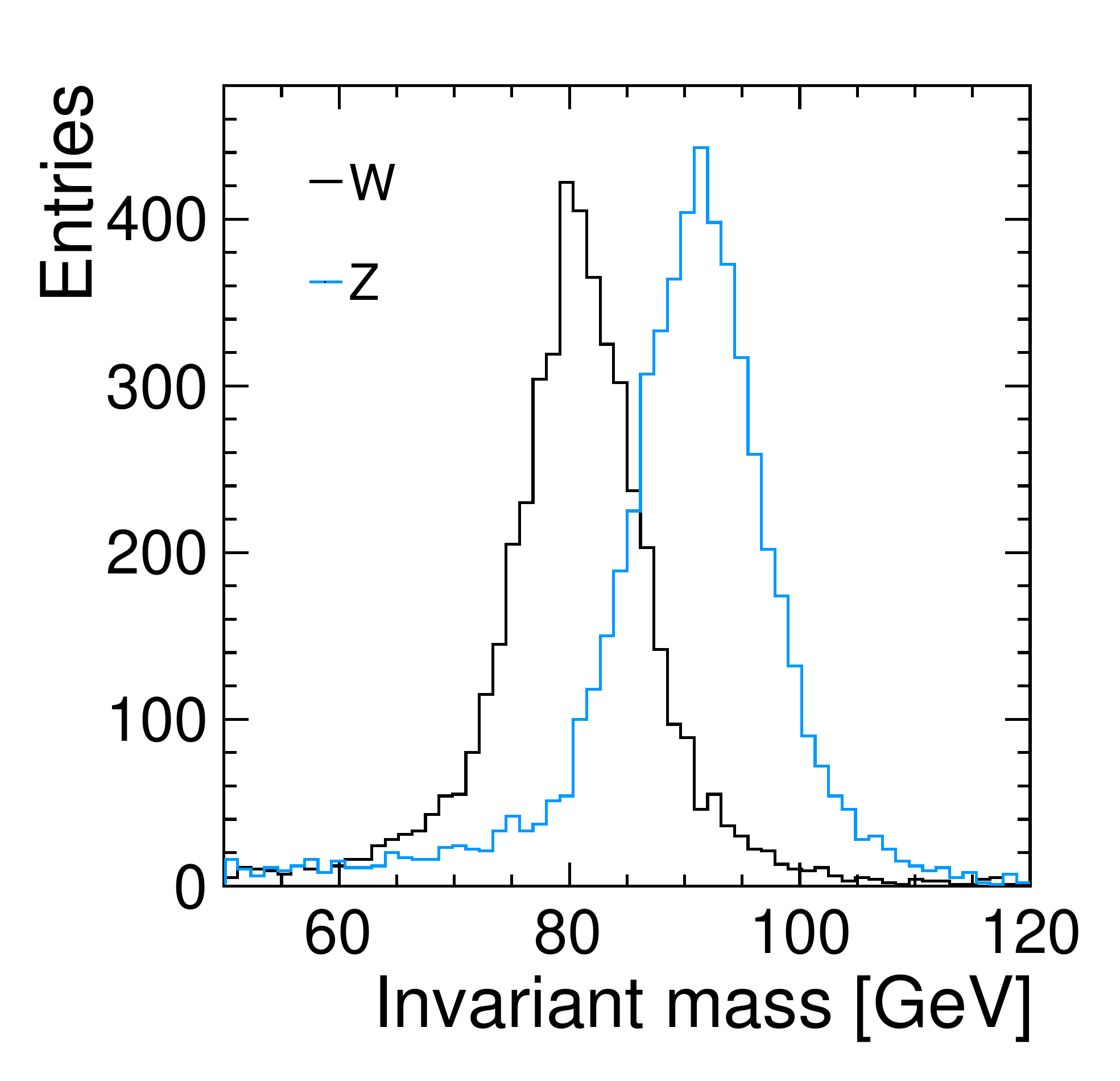}
      \caption{Mass distributions of reconstructed $\Wboson$ and $\Zzero$ at an energy of 500\,GeV, without background (left) and with 60~BX of background (right). Without background the separation is 
               2.2~$\sigma$. With 60~BX of background the separation drops to 1.8~$\sigma$. \texttt{Tight} PFO selection cuts are used for events with background.\label{figWZ_Mpeak_sep}}
    \end{center}
  \end{figure*}

  A key requirement for the physics programme at CLIC is the ability to separate hadronic $\Wboson$ and $\Zzero$ decays. 
  For this purpose, the di-jet mass distributions obtained from the simulated $\Wboson$ decays were compared to those
  obtained from $\Zzero$ decays using simulated 
  \mbox{$\epem\rightarrow\Zzero\Zzero \rightarrow \neutrino\Aneutrino \Qq\AQq$} events.
  As for the $\Wboson$ datasets, fully simulated and reconstructed events were available with $\Zzero$ energies of 125, 250, 500 and 1000\,GeV. Samples were produced without
  background and with 60~BX and 2$\times$60~BX of overlaid $\gghadrons$ background. The same reconstruction and selection procedure was used and the reconstruction performance was found to be very similar to that
  obtained for the $\Wboson$s.
  Figure~\ref{figWZ_Mpeak_sep} shows the reconstructed mass peaks for $\Wboson$ and $\Zzero$ particles with an
  energy of 500\,GeV. Without background, there is clear separation between the 
  peaks. With 60~BX of background, the separation is somewhat degraded.
   
  \begin{table*}[!ht]
    \caption{Mass resolutions obtained for reconstructed $\Wboson$ and $\Zzero$ at different energies and with different amounts of overlaid background. The separation of the $\Wboson$ and $\Zzero$ peaks
    and the identification efficiencies are labelled for each configuration. \texttt{Tight} PFO selection cuts are used for events with background. \label{tabWZ_Sep}}
    \begin{center}
      \begin{tabular}{l c c c c c }\hline
	\toprule
	BX & $E_{\Wboson,\,\Zzero}$ & $\sigma_{m(\Wboson)}/m(\Wboson)$ & $\sigma_{m(\Zzero)}/m(\Zzero)$  & Separation & $\epsilon$ \\ 
	& [GeV] & [\%]& [\%] & [$\sigma$] &[\%]\\
	\midrule
	\multirow{4}{*}{0 BX} & 125   &  4.6 & 4.2  & 2.2 & 88 \\
	& 250   & 4.3 & 4.0 & 2.2 & 89 \\
	& 500   & 4.4 & 4.2 & 2.2 & 88 \\
	& 1000  & 6.1 & 5.4 & 1.9 & 87 \\ 
	\midrule
	\multirow{4}{*}{60 BX} & 125   & 10.7 & 10.1 & 1.5 & 70\\
	& 250   &  9.3 & 9.0 & 1.6 & 74\\
	& 500   &  6.7 & 6.6 & 1.8 & 79\\
	& 1000  &  8.1 & 7.7 & 1.7 & 77\\ 
	\hline
	\multirow{4}{*}{2$\times$60 BX} & 125   & 11.8 & 11.0 & 1.4 & 67\\
	& 250   & 10.4  & 10.2 & 1.5 & 71\\
	& 500   & 7.8  & 7.9 & 1.6 & 74\\
	& 1000  & 9.1  & 8.5 & 1.6 & 73\\ 
	\bottomrule
      \end{tabular}
    \end{center}
  \end{table*}   
  
  Table~\ref{tabWZ_Sep} summarises the mass resolutions obtained for the $\Wboson$ and $\Zzero$ samples and the separation achieved and the identification efficiency for each of the 
  different energies investigated. The separation of the $\Wboson$ and $\Zzero$ peaks is quantified by determining the fraction of mis-identified events for the optimum mass cut.
  The natural widths of the $\Wboson$ and $\Zzero$ bosons restrict the identification efficiency to $<94\,\%$~\cite{ThomsonNimA}. The fraction of mis-identified events is converted into an
  equivalent Gaussian statistical separation, where for example, in the case of ideal Gaussian distributions, a mis-identification of 15.8\,\% corresponds to a separation of 2$\sigma$.
  Without background a $2\sigma$ separation of the $\Wboson$ and $\Zzero$ mass peaks is essentially maintained for gauge bosons with energies between 125\,GeV and 1\,TeV. This separation is 
  reduced to about $1.7\sigma$ when background is included.
  Even without the background, it would be exceedingly difficult to achieve such a level of separation (over such a wide range of jet energies) using a traditional approach to calorimetry.
  The successful use of particle flow calorimetry to separate hadronic $\Wboson$ and $\Zzero$ decays in a physics analysis is demonstrated in~\cite{CLIC-CDR}.

  \subsection{Measurement of Missing Momentum}
  
  The reconstruction of missing momentum is important in many physics processes. Here, the performance 
  of the particle flow reconstruction of missing momentum is quantified in two ways: i) 
  the missing momentum resolution for events with a true missing
  momentum signature; and ii) the reconstructed missing momentum in event topologies where the momentum
  of the reconstructed particles should balance.
  
  \subsubsection*{Resolution of Missing Transverse Momentum}
\begin{figure*}[!htb]
    \begin{center}
      \includegraphics[width=0.49\textwidth]{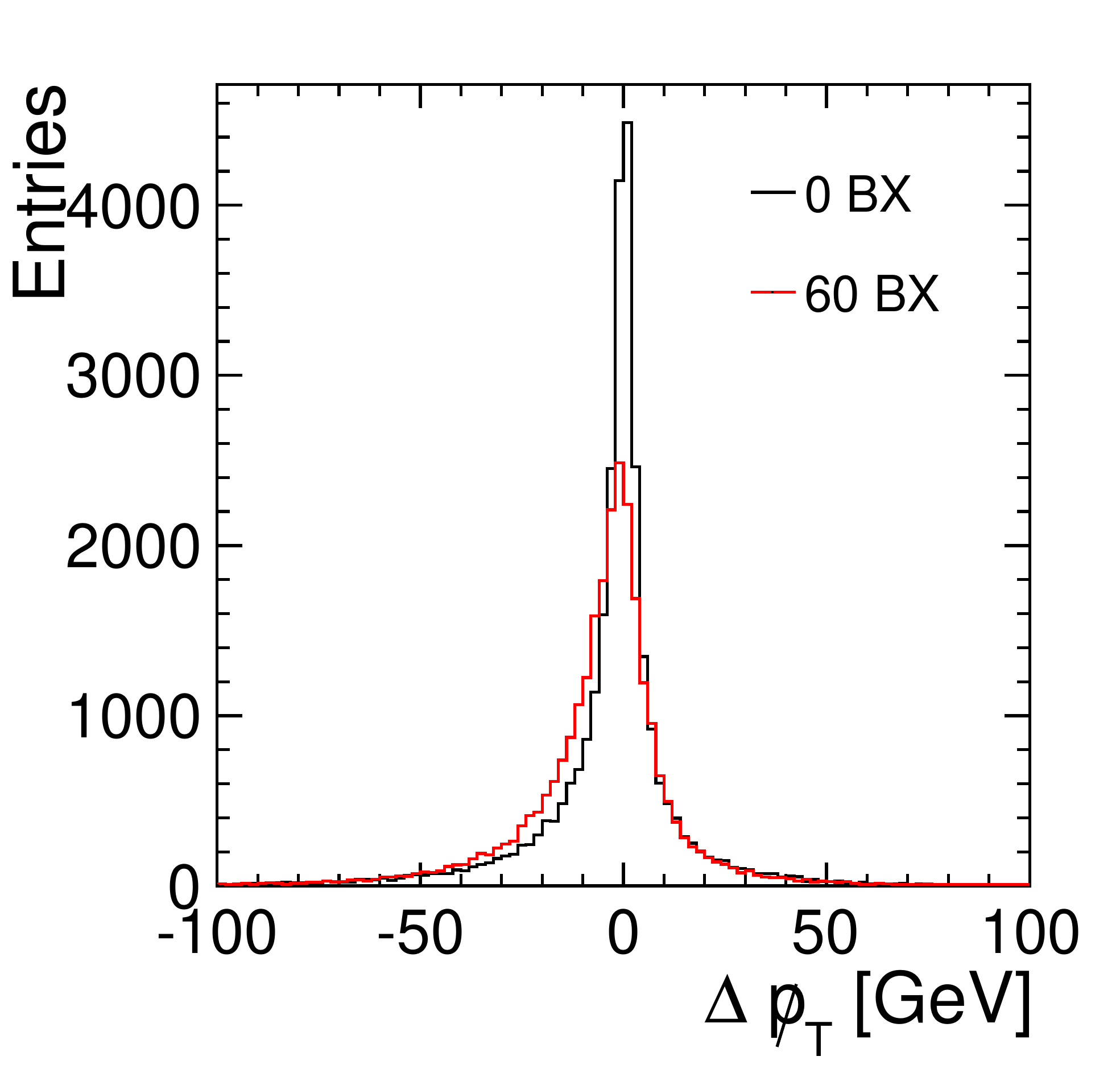}
      \includegraphics[width=0.49\textwidth]{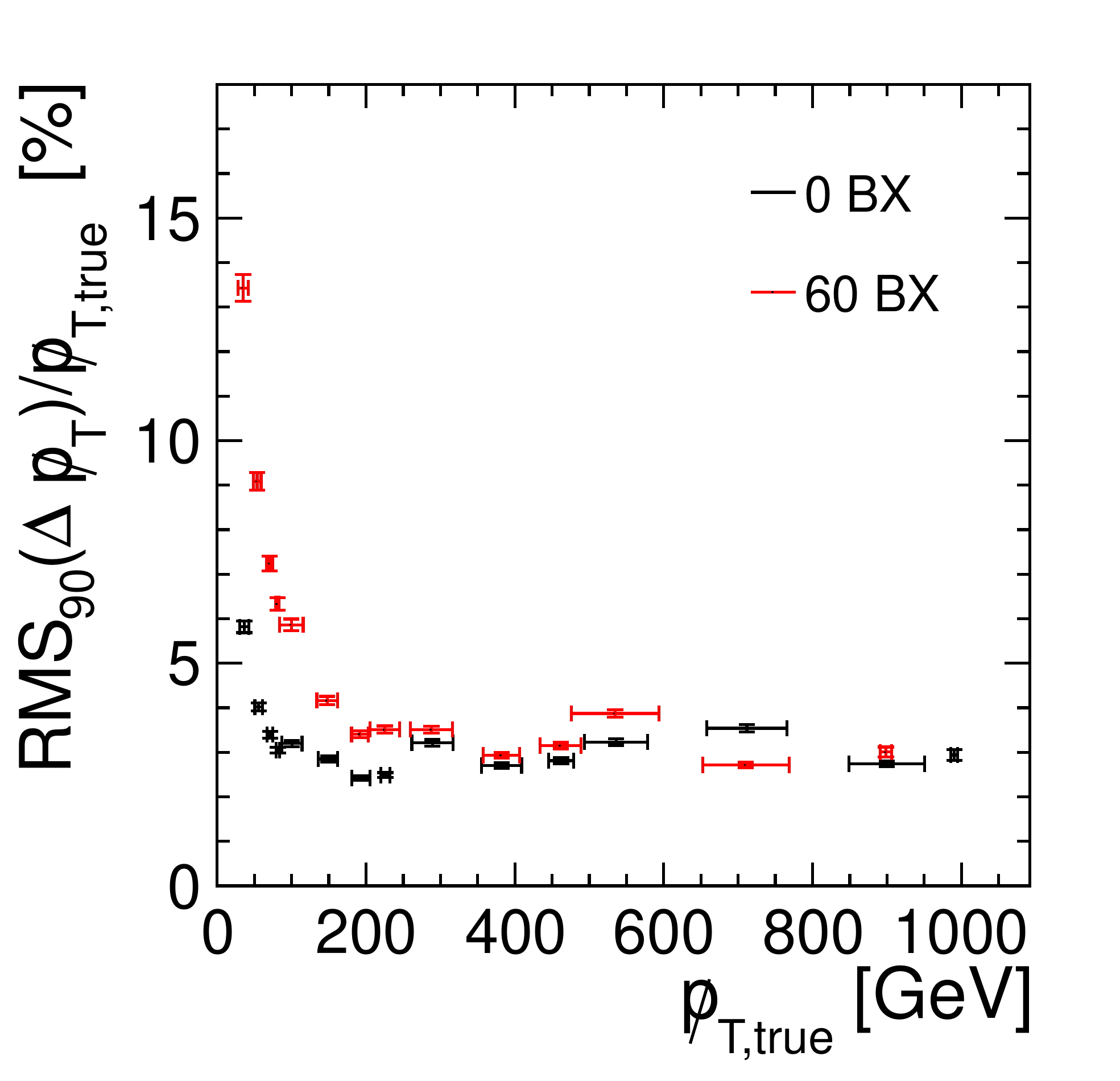}
      \caption{Distribution of the difference between measured and true missing transverse momentum (left) and the missing transverse momentum resolution (right) for 
	       \mbox{$\Zzero\Zzero\rightarrow\neutrino\Aneutrino \Qq\AQq$} events. Separate distributions are shown for events without background and with 60~BX of background. 
	       \texttt{Tight} PFO selection cuts are used for events with background. \label{figMET_distr}}
    \end{center}
  \end{figure*}

  The missing transverse momentum resolution was quantified using the 
  simulated \mbox{$\Zzero\Zzero\rightarrow\neutrino\Aneutrino \Qq\AQq$} events discussed in Section~\ref{sec:WW}. 
  The jet reconstruction was unchanged, and the only event selection requirement
  was that both jets satisfy $|\cos{\theta}| < 0.9$. The missing momentum depends on both the energies of the 
  neutrinos and their polar angles. For this reason the four different energy points, $\sqrt{s}$=250, 500, 1000 and 2000\,GeV, 
  were combined into a single data set. The missing transverse momentum, $\slashed p_{\mathrm{T}}$, 
  was calculated from the vector sum of the momenta of all the particles in the two reconstructed jets.
  This was compared to the generated missing transverse momentum of the two neutrinos, $\slashed p_{\mathrm{T,\,true}}$. 
  Figure~\ref{figMET_distr} shows the distribution of the difference between measured and true missing transverse momentum
  (for all events combined) and  the missing transverse momentum resolution as a function of the true $\pT$ in the event. 
  For each bin in $\pT$, the missing transverse energy resolution was obtained by calculating the RMS$_{90}$ of the $\Delta \slashed p_{\mathrm{T}}$ distribution and dividing by the true missing transverse
  momentum. The resolutions obtained indicate that the missing transverse momentum 
  can be measured with an asymptotic precision of about 3\,\% for $\pT>100\,\GeV$. 
  
  \subsubsection*{Fake Missing Momentum}
\begin{figure*}[!htb]
    \begin{center}
      \includegraphics[width=0.49\textwidth]{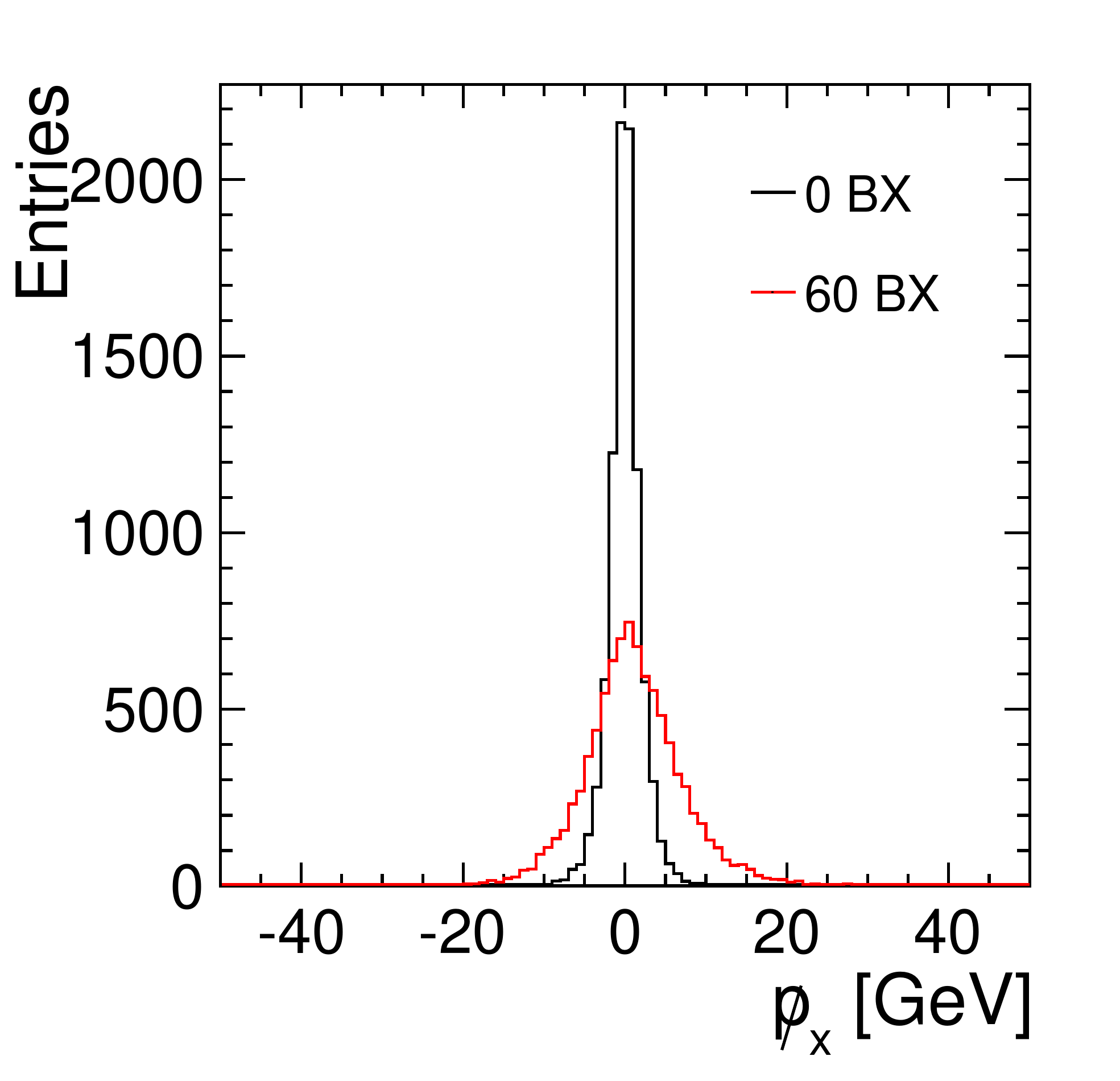}
      \includegraphics[width=0.49\textwidth]{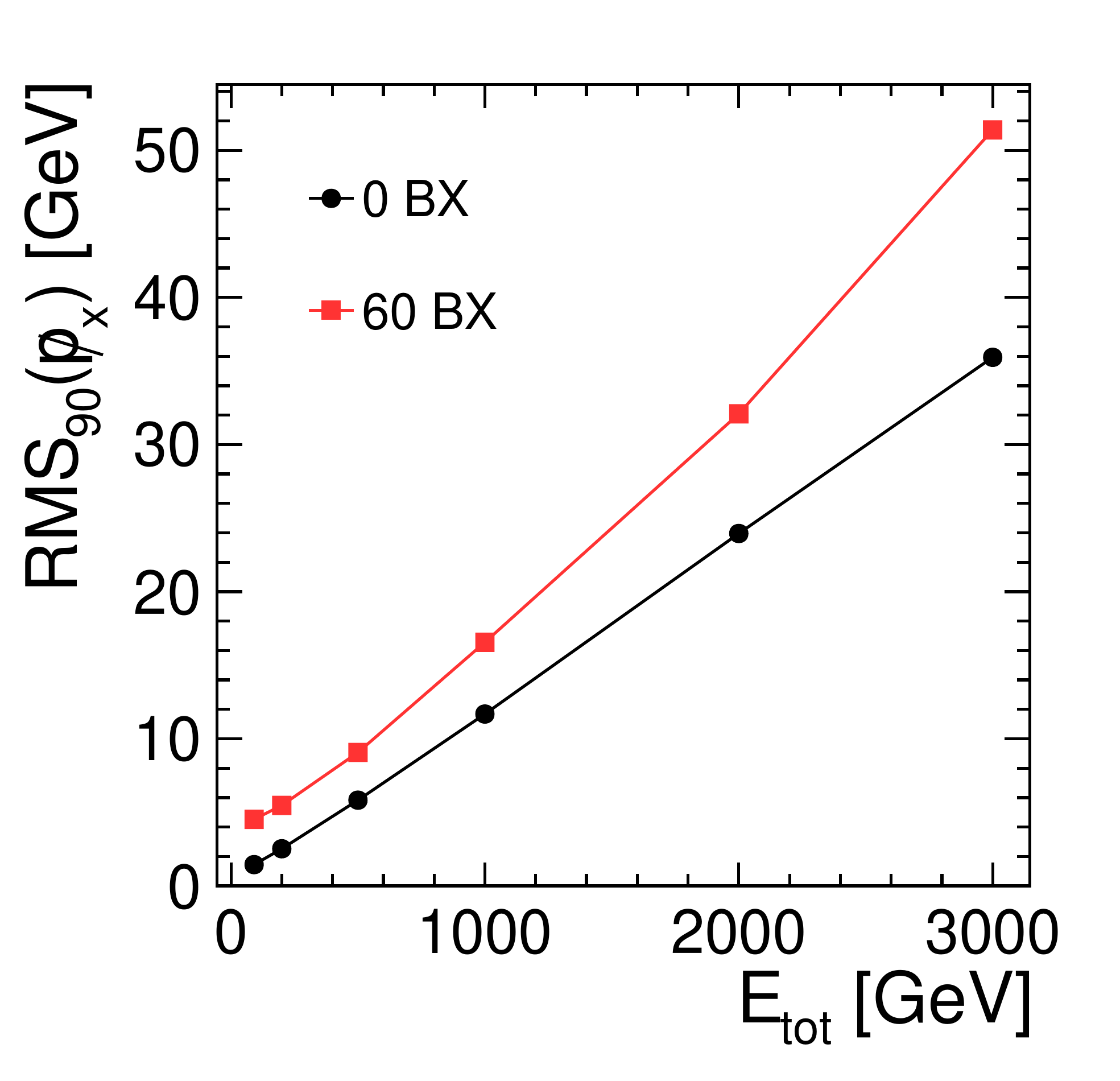}
      \caption{The $x$ component of fake missing momentum for $\Zzero^\prime$ events at 91\,GeV (left) and the RMS$_{90}$ of this distribution as a function of the total energy deposited 
               in the detector (right).Separate distributions are shown for events without background and with 60~BX of background. 
	       \texttt{Tight} PFO selection cuts are used for events with background. \label{figFMET_distr}}
    \end{center}
  \end{figure*}

  Many physics analyses rely on measurements of missing momentum. Fake missing momentum can
  result from limitations in the detector coverage and from the mis-reconstruction of the momenta of
  the particles. The fake missing momentum was quantified using the \mbox{$\Zzero^\prime\rightarrow\Qq\AQq$} events
  described in Section~\ref{sec:TechnicalStudy}.
  The events were reconstructed into three jets, to account for the possibility of a single hard gluon radiation. 
  Figure~\ref{figFMET_distr} shows the distribution of the $x$ component of the fake missing momentum for the 91\,GeV $\Zzero^\prime$ sample. The resolution was quantified by calculating the 
  RMS$_{90}$ of this distribution.
  The right hand plot of Figure~\ref{figFMET_distr} shows that the RMS$_{90}$ rises approximately linearly with the total energy deposited in the detector.
  Overall, the level of fake missing transverse momentum (in one coordinate) is at the level of 1-2\,\% of the event energy. The impact of the $\gghadrons$ background is not large for the event 
  energies of interest at CLIC. 
  
  \section{Summary}
  Particle flow calorimetry is an extremely powerful technique, which can deliver unprecedented energy and mass resolutions at future linear colliders, even in the challenging environment at CLIC. The
  PandoraPFA particle flow algorithm now offers an accurate particle flow reconstruction at jet energies from 50\,GeV to 1.5\,TeV, 
  and includes a managed transition to energy flow calorimetry if calorimeter occupancies become exceptionally high. Following the particle flow reconstruction, timing cuts can be applied to reconstructed 
  particles to significantly reduce the contribution of beam-induced backgrounds.

  The energy dependence of the particle flow reconstruction, and the systematic effects of detector acceptance, have been investigated via a study of mono-energetic jets with energies up to 1.5\,TeV. 
  Two physics channels, providing $\Wboson$ and $\Zzero$ particles, were then chosen to study the performance of particle flow calorimetry with jet reconstruction and in the presence of background. The energy 
  and mass resolutions achieved are very promising. At higher energies, the effect of the $\gghadrons$ background on the energy resolution is negligible. The mass resolution is more sensitive to the 
  quality of the reconstructed jets, and so is more strongly affected by backgrounds, but the separation of hadronic $\Wboson$ and $\Zzero$ decays remains achievable.
  Even though PandoraPFA is not optimised for the reconstruction of missing momentum, it allows missing transverse momentum to be measured with the same precision as jet energies. The level of fake
  missing transverse momentum (in one coordinate) is at the level of 1-2\,\% of the event energy. 
  
  Compared to previous results, obtained in~\cite{ThomsonNimA}, significant perfomance improvements have been achieved with the PandoraPFA algorithms, especially for higher jet energies. These improvements
  are evident in the jet energy resolution and in the ability to distinguish between the hadronic decays of $\Wboson$ and $\Zzero$ particles. The PandoraPFA reconstruction has been used for all the simulated 
  physics studies presented in the CLIC Conceptual Design Report~\cite{CLIC-CDR}. 

  \section*{Acknowledgements}
  This work was funded in part by the UK Science and Technology and Facilities Council and by the European Union under the Advanced European Infrastructures for Detectors and Accelerators (AIDA) project.

\end{document}